\documentclass[12pt,preprint]{aastex}
\begin{document}

\title{Discovery of kHz QPOs and shifted frequency correlations in the
accreting millisecond pulsar XTE~J1807--294}

\author{Manuel Linares\altaffilmark{1}, Michiel van der Klis\altaffilmark{1}, Diego Altamirano\altaffilmark{1}, Craig B. Markwardt\altaffilmark{2,3}}

%\date{July 20th, 2005}

\def\rem#1{{\bf (#1)}}
\def\hide#1{}

\altaffiltext{1}{Astronomical Institute ``Anton Pannekoek,'' University of Amsterdam and Center for High-Energy Astrophysics, Kruislaan 403, NL-1098 SJ Amsterdam, Netherlands.}
\altaffiltext{2}{Laboratory for High Energy Astrophysics, NASA Goddard Space Flight Center, Greenbelt, MD 20771.}
\altaffiltext{3}{Department of Astronomy, University of Maryland, College Park, MD 20742.}

%\keywords{}

\begin{abstract}

We report the discovery of twin kHz QPOs in the X ray flux of 
XTE~J1807--294, the fourth accreting millisecond pulsar (AMP).  This is
the second AMP exhibiting twin kHz QPOs. In contrast to the first
case, SAX J1808.4--3658, the frequency separation $\Delta\nu$ between
the kHz QPOs is consistent with the pulse frequency (190.6~Hz), not
with half that value, confirming for the first time from pulsation
measurements the inference, based on burst oscillations, that 'slow
rotators' (spin frequency less than 400~Hz) have $\Delta\nu$
approximately equal to the spin frequency.  While the QPOs move in
frequency together over a range of more than 200~Hz, $\Delta\nu$
remains constant with an average value of 205$\pm$6~Hz. Variability
components were found in the 5--130~Hz range similar to those seen in
other LMXBs. The correlations between the QPO and noise frequencies
are also similar to those in other sources, but shifted by a factor of
1.59 in kHz QPO frequencies, similar to the factor 1.45 shift found
for SAX~J1808.4--3658. Our results argue in favour of a
spin-related formation mechanism for twin kHz QPOs and against a
spin-related cause of the shift in the frequency correlations.
\end{abstract}

\maketitle

\section{Introduction}
\label{sec:intro}

The accretion-driven millisecond X-ray pulsar (AMP) family is
gradually growing. Since the discovery of its first member \citep[SAX
J1808.4--3658,][]{Wijnands98} six more have been added: XTE
J1751--305, XTE~J0929--314, XTE~J1807--294, XTE~J1814--338 and
recently IGR~J00291+5934 and HETE~J1900.1--2455 \citep[respectively:][]{Markwardt02,
Remillard02, Markwardt03a, Markwardt03b, Galloway05, Vanderspek05}. All of these are
in transient low-mass X-ray binaries (LMXBs) showing faint outbursts
(peak count rate 10--100 mCrab) every few years that last a few
weeks. Their pulse frequency (i.e., the neutron-star spin frequency)
lies in the range 185--599~Hz and the orbital period is between 40
minutes and 2.5 hours. See \citet{Wijnands05} for an updated
observational review.

XTE~J1807--294 is the fourth-discovered accreting millisecond pulsar.
The source was detected first on February $13^{th}$, 2003, in periodic 
scans of the Galactic bulge region using the Proportional Counter Array 
(PCA) instrument on board the Rossi X-ray Timing Explorer (RXTE). These 
scanning observations show a rising flux, with the epoch of maximum X-ray 
flux (count rate of $\sim 58$ mCrab) occurring between February $19^{th}$ and $22^{nd}$ \citep{Markwardt03c}. 
On February $21^{st}$, 2003, in the first pointed observation of the source 
by the PCA, 190.6 Hz pulsations were detected, confirming this source as an 
accreting millisecond pulsar \citep{Markwardt03c}. An orbital period of
$\sim$40 minutes was determined by \citet{Markwardt03a}, still the shortest 
among AMPs. A pulse analysis of the outburst was performed by \citet{Markwardt04}.

XTE~J1807--294 had the longest outburst among the AMPs observed up to now.  It
was followed with the PCA for
five months, although after a few weeks the source became too weak to
measure its aperiodic variability. As noted in \citet{Falanga05}, the
X-ray flux decay did not show a break to a faster decline, such
as in the other AMPs \citep{Wijnands05} and attributed in those
objects to the propeller effect \citep[e.g.,][]{Gilfanov98}.
\citet{Falanga05} described the energy spectrum as a sum of a disk
blackbody and a thermal Comptonization component, without evidence of
spectral lines or reflection. No optical or infrared counterpart has
been reported to date. Assuming a distance of 8kpc, \citet{Campana05} placed an upper limit of
4$\times10^{31}$ erg $s^{-1}$ on the unabsorbed 0.5--10 keV quiescent
luminosity. 

Studying the power spectra of LMXBs is one of the most direct ways of
analyzing the accretion flow in these systems. In particular the
high-frequency power-spectral features constitute a probe of the
innermost regions of the accretion disk, where gravitational fields
are extreme \citep[see][for reviews]{vanderKlis00, vanderKlis04a}. The
first discovered AMP, SAX~J1808.4--3658 \citep{Wijnands98}, in its
2002 outburst provided surprising information on the relation between
the neutron-star spin and the kHz QPOs and confirmed expectations
concerning the burst oscillations. While the burst oscillations \citep[cf.][]{Strohmayer96}
occurred at the pulse, and inferred spin, frequency 
\citep{Chakrabarty03}, the separation $\Delta\nu$ between the kHz QPO
frequencies was consistent with {\it half} the pulse frequency
\citep{Wijnands03}.  Although based on only a single simultaneous
detection of twin kHz QPOs, this result carried important consequences
for our understanding of LMXBs, as it falsifies simple spin-orbit
beat-frequency models, yet requires an intimate link between spin and
kHz QPOs.  This led to several new proposed models for the nature of
the interaction between disk flow and neutron-star spin
\citep{Wijnands03, Lamb03, Kluzniak04, Lee04} replacing, modifying or adding to earlier proposed models
(e.g., \citealt{Miller98, Stella98, Abramowicz03}). So far, all
sources with inferred spin frequencies $\nu_{spin}>400$~Hz ('fast
rotators') have $\Delta\nu\approx\nu_{spin}/2$ and the ones with
$\nu_{spin}<400$~Hz ('slow rotators') have $\Delta\nu\approx\nu_{spin}$.

Several correlations have been found between the frequencies of
power-spectral components in Z and atoll sources \citep[the two main types
of neutron-star LMXBs; ][]{Hasinger89}, as well as among black hole
candidate systems.  Initially, the WK relation \citep{WK99} and the
PBK relation \citep{PBK99} were identified.  Including several more
correlated components, this set of correlations was synthesized into a
'universal scheme of correlations' among atoll sources by
\citet{Straaten03}.  Comparing AMPs to other atoll sources,
\citet{Straaten05} found that the correlations in two of the AMPs are
shifted relative to those of the other atoll sources.  The simplest
description of the shifts was that the two kHz QPOs had frequencies a
factor $\sim 1.5$ lower than in the other sources.

In this paper we analyze all the RXTE PCA data on XTE~J1807--294, and
report the discovery of pairs of simultaneous kHz QPOs.  The
separation $\Delta\nu$ between the seven pairs of kHz QPOs we detect
is consistent with the pulse frequency, and confirms for the
first time from pulsations that the slow rotators have a kHz QPO
frequency separation that is approximately equal to the neutron star
spin frequency.  Finally, including also the low-frequency variability components 
in our study, we find a shift in the frequency correlations by a factor $\sim$1.5, 
similar to that seen in SAX~J1808.4--3658.

\section{Observations and Data Analysis}
\label{sec:data}

We used all the RXTE PCA data from XTE~J1807--294 available in NASA's
High Energy Astrophysics Science Archive Research Center
taken between February $27^{th}$ and July $29^{th}$, 2003.

For the color analysis we used the Standard 2 data, which have a 16-s
time resolution and 129 energy channels. We divided the channels into
four bands in a way that is often used for neutron-star sources: band
A: 2.0--3.5 keV, band B: 3.5--6.0 keV, band C: 6.0--9.7 keV, band D:
9.7--16.0 keV. The count rates in these precise bands were estimated
by interpolating linearly between the corresponding energy
channels. The background was estimated using the standard bright source
background models. From the counts in each band, we
computed the colors and intensity as: hard color=D/C, soft color=B/A,
intensity=A+B+C+D. We then normalized colors and intensity to the Crab values nearest
in time \citep{Kuulkers94} and in the same PCA gain epoch (e.g.,
\citealt{Straaten03}).

The timing analysis was done using the GoodXenon data, with the
original $1\mu s$ time resolution data rebinned into 1/8192-s bins, and
including all 256 energy channels.  We performed fast
Fourier transforms (FFTs) of 128-s data segments, fixing the frequency
resolution and the lowest available frequency to $\sim$0.008~Hz; the
highest available Fourier frequency (Nyquist frequency) was 4096 Hz. No background subtraction or dead-time correction was made prior to the FFTs.
The resulting power spectra were ``weeded'' (the pulsar spike at
$\sim$ 190~Hz was removed at full frequency resolution) and a Poisson
level was subtracted. Following \citet{Kleinwolt04} we first estimated
the Poisson noise using the \citet{Zhang95} formula and then (after
inspecting the $\sim$3000--4000~Hz range and finding no unexpected
features) shifted it to match the level between 3072--4096~Hz,
where no intrinsic power should be present, but only counting
statistics noise (this shift was in all cases smaller than
0.15\% of the previously estimated Poisson level). Then we normalized 
the power spectra using the rms normalization \citep{vanderklis95b}. 
Several observations with similar power spectra
and colors were averaged into sets labeled A--H in order to improve
the statistics (see Table~\ref{table:obs}). The sets were ordered by
increasing kHz QPO frequencies.

To fit the overall broad-band power spectra we use a fit function
consisting of the sum of several Lorentzians in the so-called
``$\nu_{max}$ representation'', described in \citet{Belloni02}. In
this representation, if $\nu_0$ is the Lorentzian's centroid frequency
and $\Delta$ its HWHM (half width at half maximum), $\nu_{max}=\sqrt{\nu_0^2+\Delta^2}$ gives the
characteristic frequency of the feature (near the centroid if it is
narrow and near the half-width if it is wide). The quality factor
$Q=\nu_0/2\Delta$ is a measure for the coherence of the variability
feature. Its strength is given by the integral power whose square
root, in the normalization we use, is the fractional rms amplitude of
the variability.  Four to five Lorentzian components were needed, all
of them significant to more than 3$\sigma$, with one exception noted
in Table~\ref{table:psfits}. In some cases in order to avoid a meaningless 
negative coherence, $Q$ was fixed to zero, which is equivalent to fitting a
zero-centered ($\nu_0=0$) Lorentzian. From now on, consistent with
previous work \citep[e.g.,][]{Belloni02, Straaten03, vanderKlis04a} we
refer to these components as $L_i$, where the L stands for Lorentzian
and '$i$' is the label identifying the component ('$b$' for break,
'$u$' for upper kHz, '$\ell$' for lower kHz, '$h$' for hump, '$LF$'
for low frequency QPO). Following this notation we call $L_i$'s
characteristic frequency $\nu_i$ and its coherence $Q_i$.  In order to
measure the kHz QPO centroid frequencies and their frequency
difference $\Delta\nu$, we performed exactly the same fits to the same
data sets, now representing the
Lorentzians' frequency by their centroid frequency $\nu_{i,0}$. While of course the two
representations are mathematically equivalent, refitting was necessary
to calculate the correct error bars for the frequencies. We split the sets 
in the original observations (see Table~\ref{table:obs}) and checked wether this led to a better 
measurement of $\Delta\nu$. Only in one case the statistical significance 
of the kHz QPOs increased (obs. 80145-01-03-00 in set H; $\Delta\nu$=197.8$\pm$7.6 
was closer to $\nu_{spin}$) but in the final analysis we conservatively used the original 
datasets. Hereafter, when we talk about frequencies this refers always to characteristic
frequencies ($\nu_{max}$) as given above; whenever centroid
frequencies ($\nu_0$) are intended this is explicitly stated.

Power law functions were fitted to the $\nu_i$ vs. $\nu_u$ relations
(where '$i$' is one of the labels mentioned above) in order to
quantify the correlations between these frequencies.  For this, we
used the FITEXY routine of \citet{Numerical} that takes into account
errors in both coordinates and fitted straight lines to the $\log\nu_i$
vs. $\log\nu_u$ relations.

To quantify the shifts in the XTE~J1807--294 correlations and compare them
with the shifts in the other AMPs we used the same method as
\citet{Straaten05}: we took as a reference the atoll sources
4U~0614+09, 4U~1608--52, 4U~1728--34 and Aql~X-1, and included only the
frequency sets with $\nu_u<600$~Hz (as noted in \citet{Straaten05}, the 
behavior of the power spectral components above that value becomes more 
complex; see also Section~\ref{sec:discussion}). We multiplied the
upper kHz QPO frequencies of XTE~J1807--294 by a factor $f$, which we
varied between 0.01 and 5 with steps of 0.001. In each step we fitted
a power law to the XTE~J1807--294 and atoll source points together.
The step with the best $\chi^2$ gives us the optimal shift factor. 
 We also calculated the shift factors
in $\nu_i$, following exactly the same procedure, but multiplying the
$\nu_i$ instead of $\nu_u$ by a factor $f$.  In the above mentioned
atoll sources there were only very few detections of $L_{LF}$.
Therefore, to obtain some information about the shift in this
component, we took as a reference the $L_{LF1}$ component of the low
luminosity bursters 1E 1724--3045 and GS 1826--24 \citep{Straaten05},
and again applied the same procedure.

\section{Results}
\label{sec:results}

\begin{figure}
\center
%\resizebox{0.7\columnwidth}{!}{\rotatebox{0}{
\includegraphics[scale=.7]{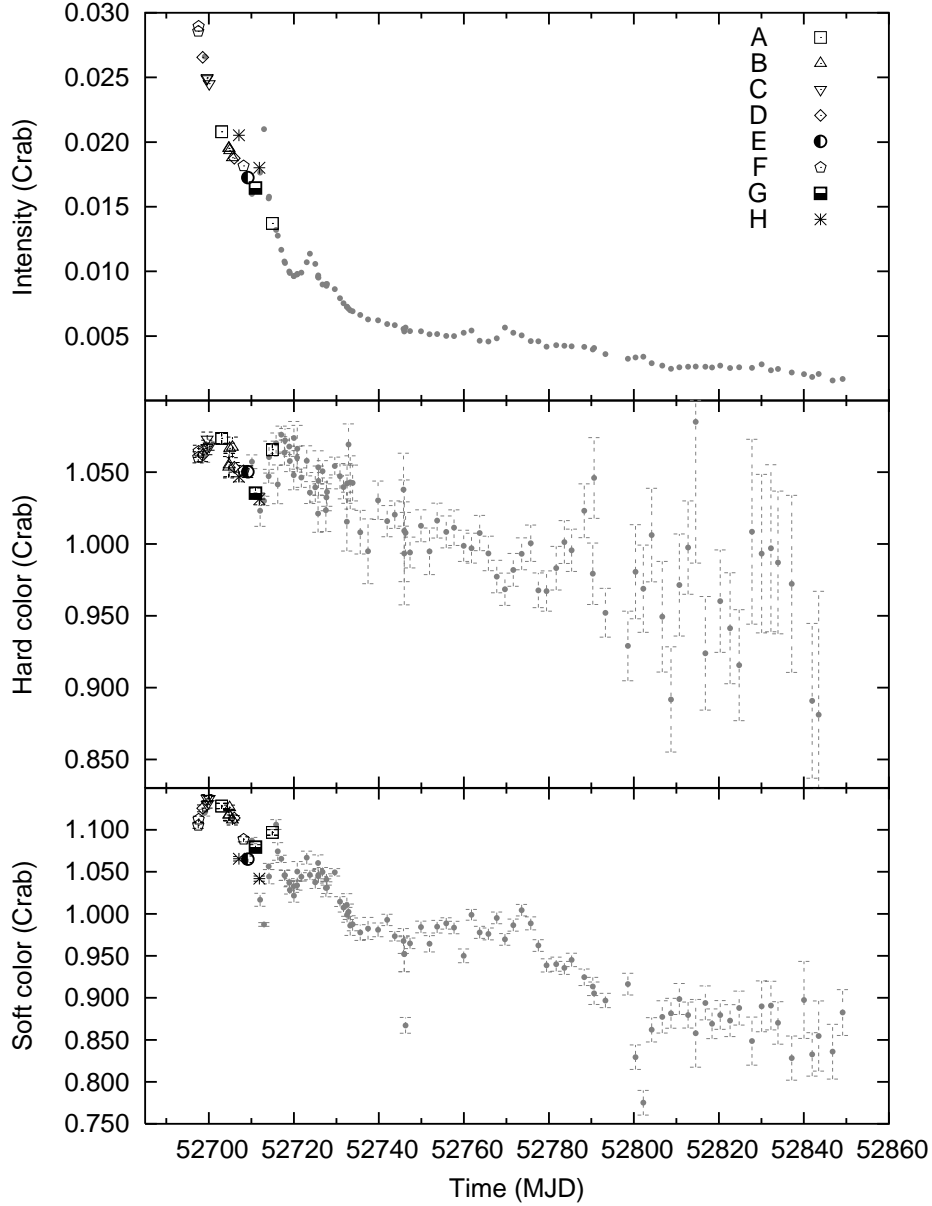}
%}}
\caption{Intensity, hard color and soft color (defined in text) versus
time for all the observations.  Symbols identify
the observations used for the timing analysis as indicated. The peak count rate ($\sim 58$ mCrab) occurred near February 21$^{st}$ (MJD 52691). No strong timing features 
were detected after March 23$^{th}$ (MJD 52721) because the count rate was too low.}
\label{fig:timecol}
\end{figure}

\begin{figure}
\center
%\resizebox{1\columnwidth}{!}{\rotatebox{0}{
\includegraphics[scale=.7]{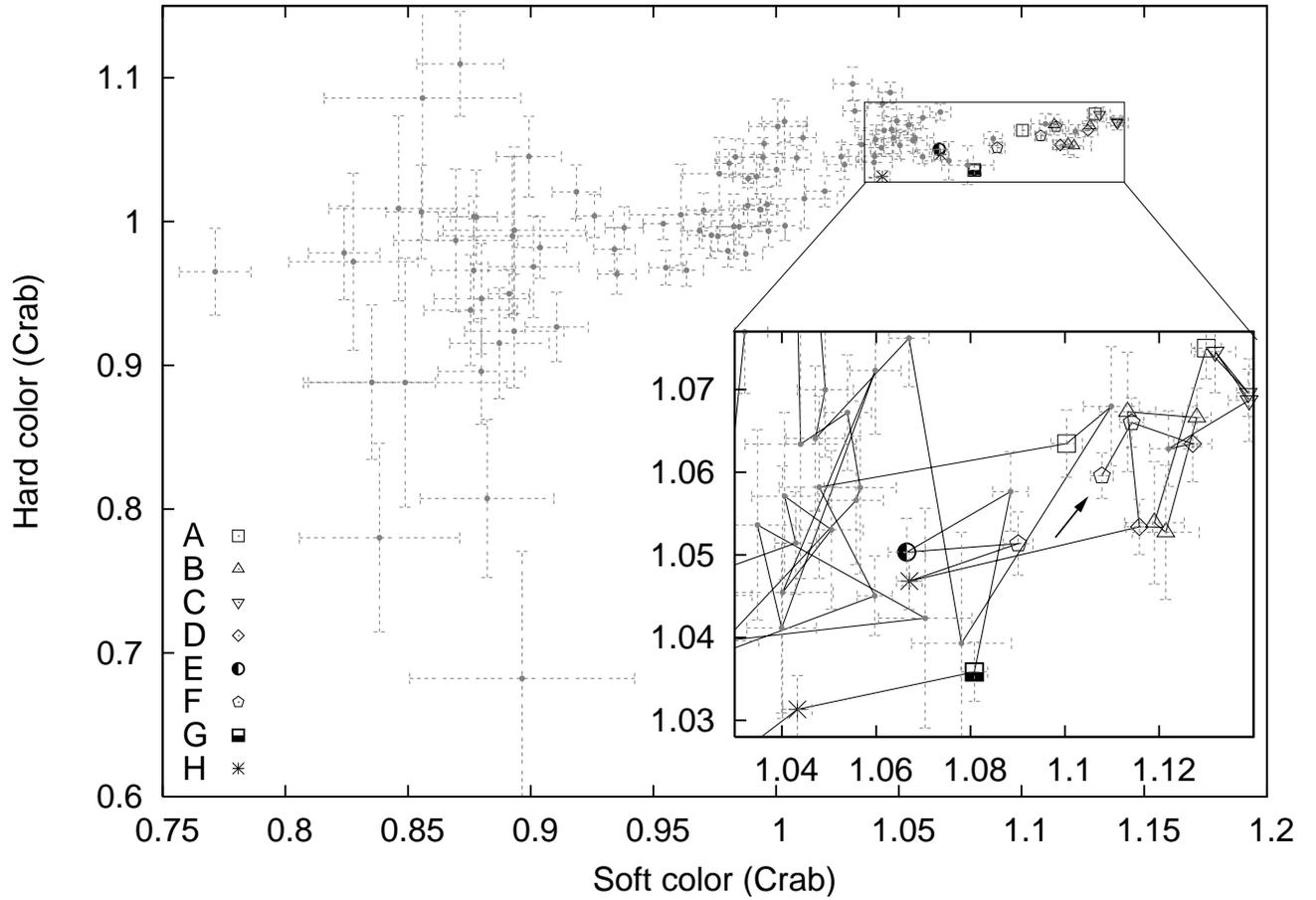}
%}}
\caption{Color-color diagram with a zoom into the area used for timing. Symbols identify
the observations used for the timing analysis as indicated.  The arrow
points at the first observation and the solid line indicates the path followed
in time.}
\label{fig:cd2}
\end{figure}

\begin{figure}
\center
%\resizebox{0.7\columnwidth}{!}{\rotatebox{0}{
\includegraphics{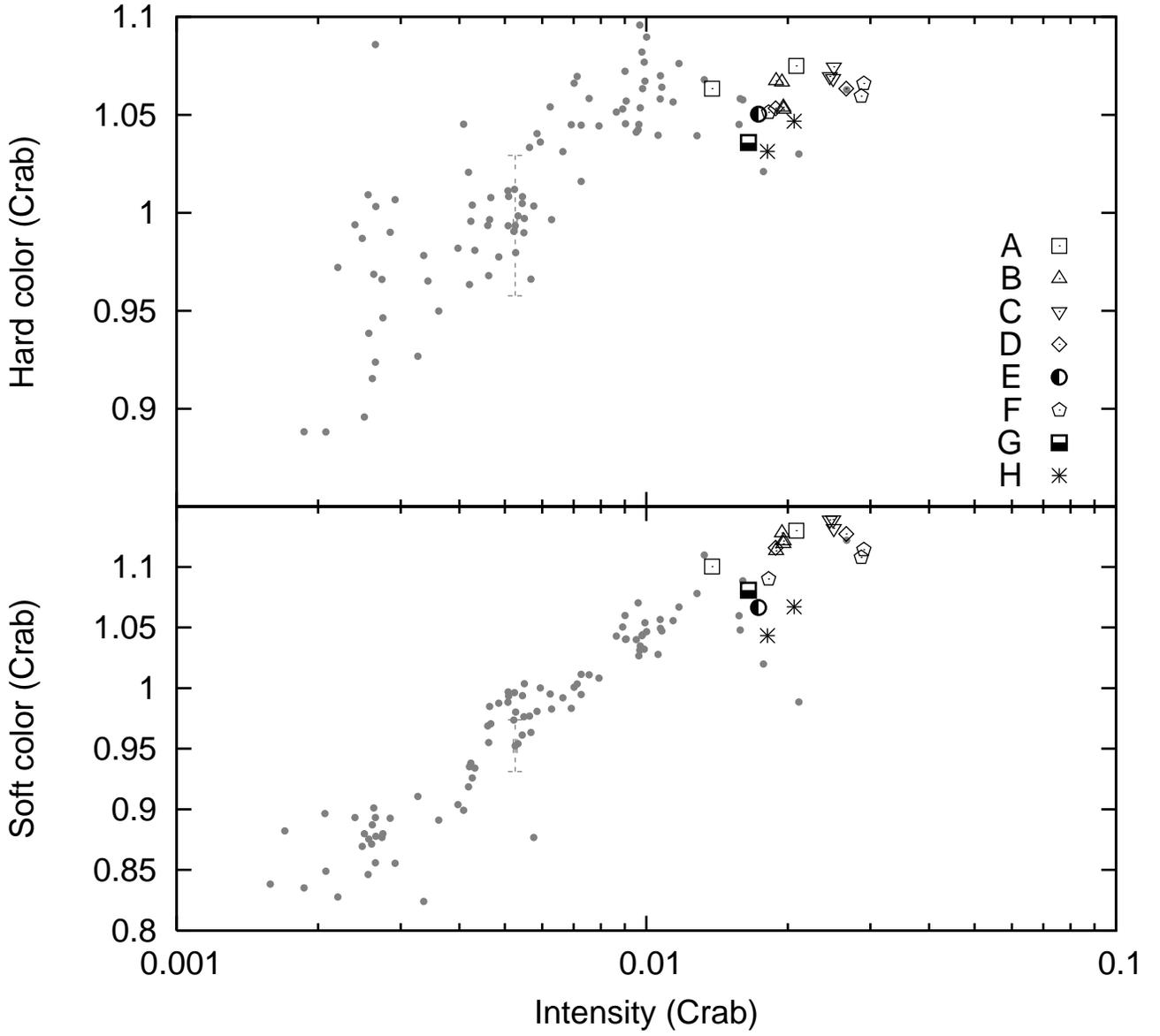}
%}}
\caption{Color intensity diagrams. Symbols identify 
the observations used for the timing analysis as indicated. Typical
error bars are shown.}
\label{fig:cd}
\end{figure}

Fig.~\ref{fig:timecol} shows the light curve for all the observations, as well as
the evolution in time of the soft and hard colors (for definitions see
Section \ref{sec:data}). As noted in \citet{Falanga05} three peaks in
the intensity occurred on March the 9$^{th}$, 14$^{th}$ and 25$^{th}$;
a peak on March the 30$^{th}$ was instrumental (caused by a flare in PCU4) and removed in the current
analysis.  The overall trend in the colors is a gradual decline
following a slight increase during the first two days of
observations. The color-color and color-intensity diagrams are shown
in Figs.~\ref{fig:cd2} and \ref{fig:cd}, respectively. The region of
the color diagrams spanned by the observations, as well as the overall
shape of the power spectra (see below), is reminiscent of the island
state in atoll sources \citep[see][]{vanderKlis04a}. As can be seen from 
Figs.~\ref{fig:timecol}-\ref{fig:cd}, there is a tendency for the colors 
to decrease with increasing QPO frequencies (from set A to set H), 
but the correspondence is not monotonic.

\begin{figure}
\center
%\resizebox{0.7\columnwidth}{!}{\rotatebox{0}{
\includegraphics{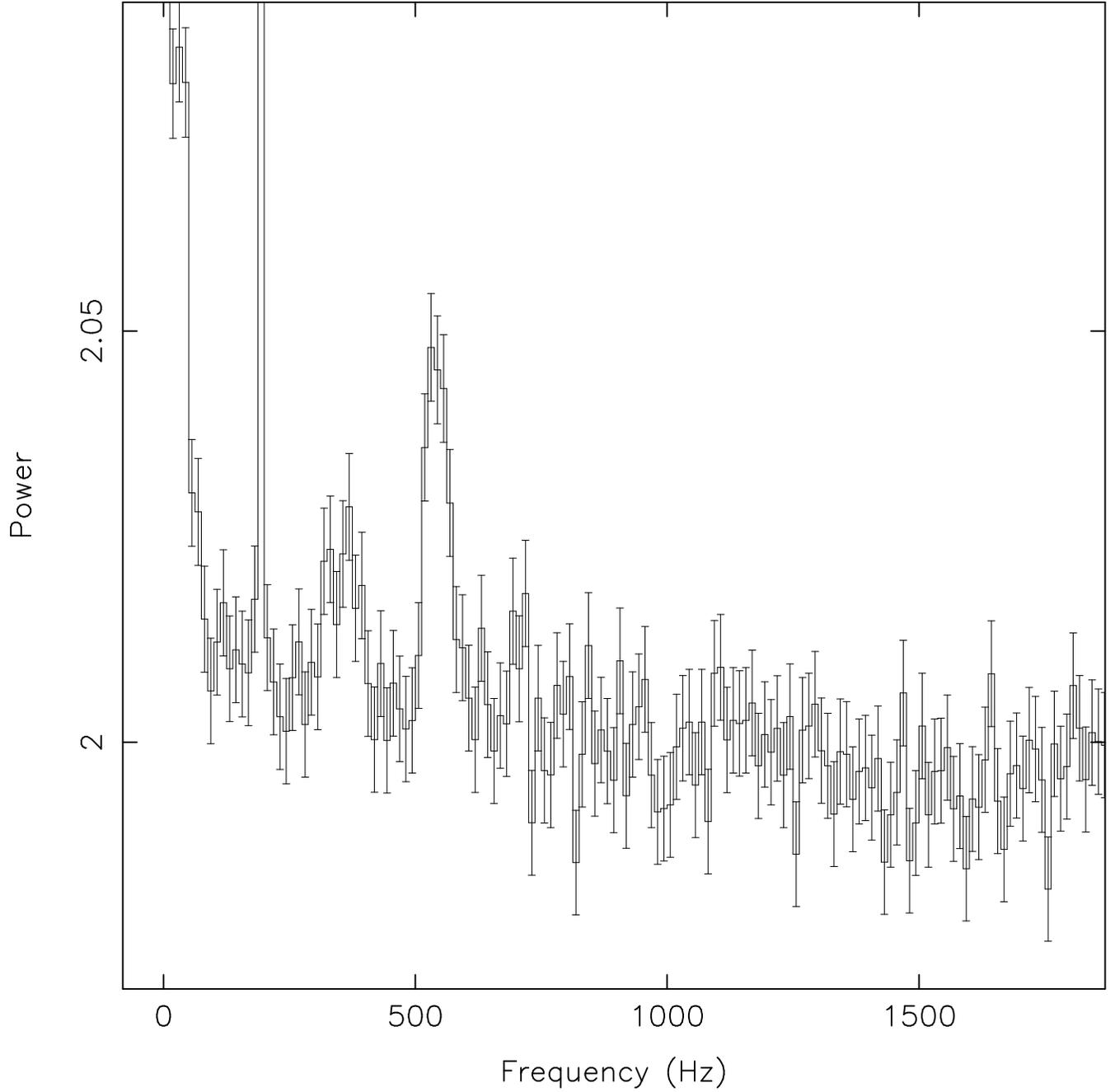}
%}}
\caption{Twin kHz QPOs in the power specrum of obs. 80145-01-03-00 (in dataset H). Note the pulsar spike just below 200 Hz.}
\label{fig:qpo}
\end{figure}

\begin{figure}
\center
%\resizebox{1\columnwidth}{!}{\rotatebox{-90}{
\includegraphics[scale=.7,angle=-90]{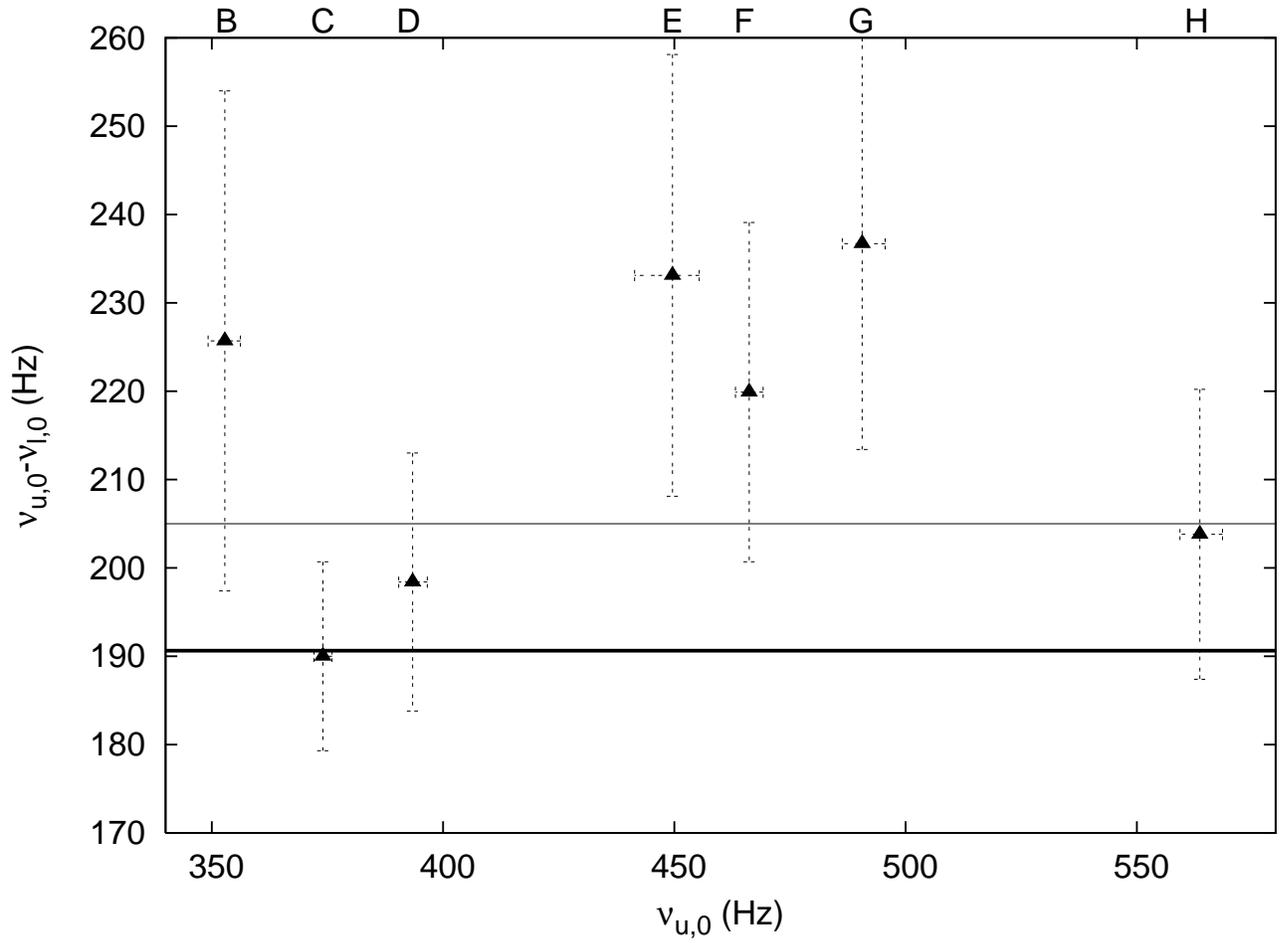}
%}}
\caption{kHz QPO frequency separation $\Delta\nu$ plotted
vs.upper kHz QPO centroid frequency $\nu_{u,0}$. The thick line indicates the pulse
frequency, the thin line the average separation and the letters above the frame the datasets used.}
\label{fig:deltanu}
\end{figure}

Two simultaneous kHz QPOs (see Fig.~\ref{fig:qpo}) were detected in
seven of the eight datasets used in this paper; the best-fit centroid
frequencies of these QPOs are listed in Table~\ref{table:qpos}
together with centroid frequency differences and ratios.  As discussed
below, the identity of the Lorentzian below $L_u$ is not clear in set
A as it could be fitting either $L_{\ell}$ or $L_{hHz}$, or both, so
we did not include this value in the analysis of the kHz QPO centroid
frequencies.  The measured differences $\Delta\nu$ between the kHz QPO
centroid frequencies are consistent with being equal to the pulse, and
inferred spin, frequency of 190.6~Hz (see Fig.~\ref{fig:deltanu}),
with no evidence for variations over the range the QPOs were detected.
The weighted average of the seven values for $\Delta\nu$ was
205.1$\pm$6.4~Hz, slightly (2.3$\sigma$) higher than the pulse
frequency.

\begin{figure}
%  \begin{center}
%  \resizebox{0.33\columnwidth}{!}{\rotatebox{0}{
\epsscale{.33}
\plotone{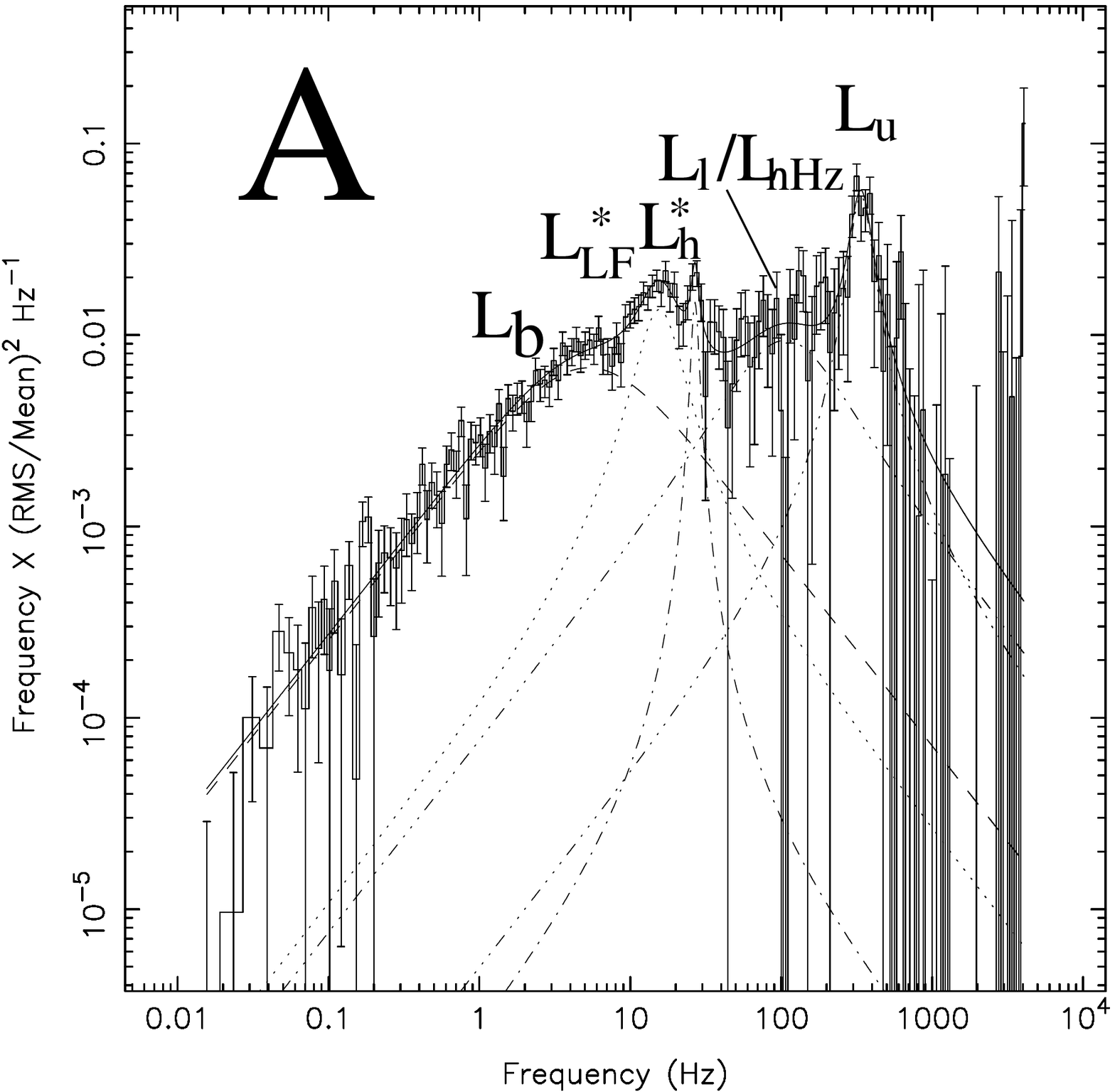}\plotone{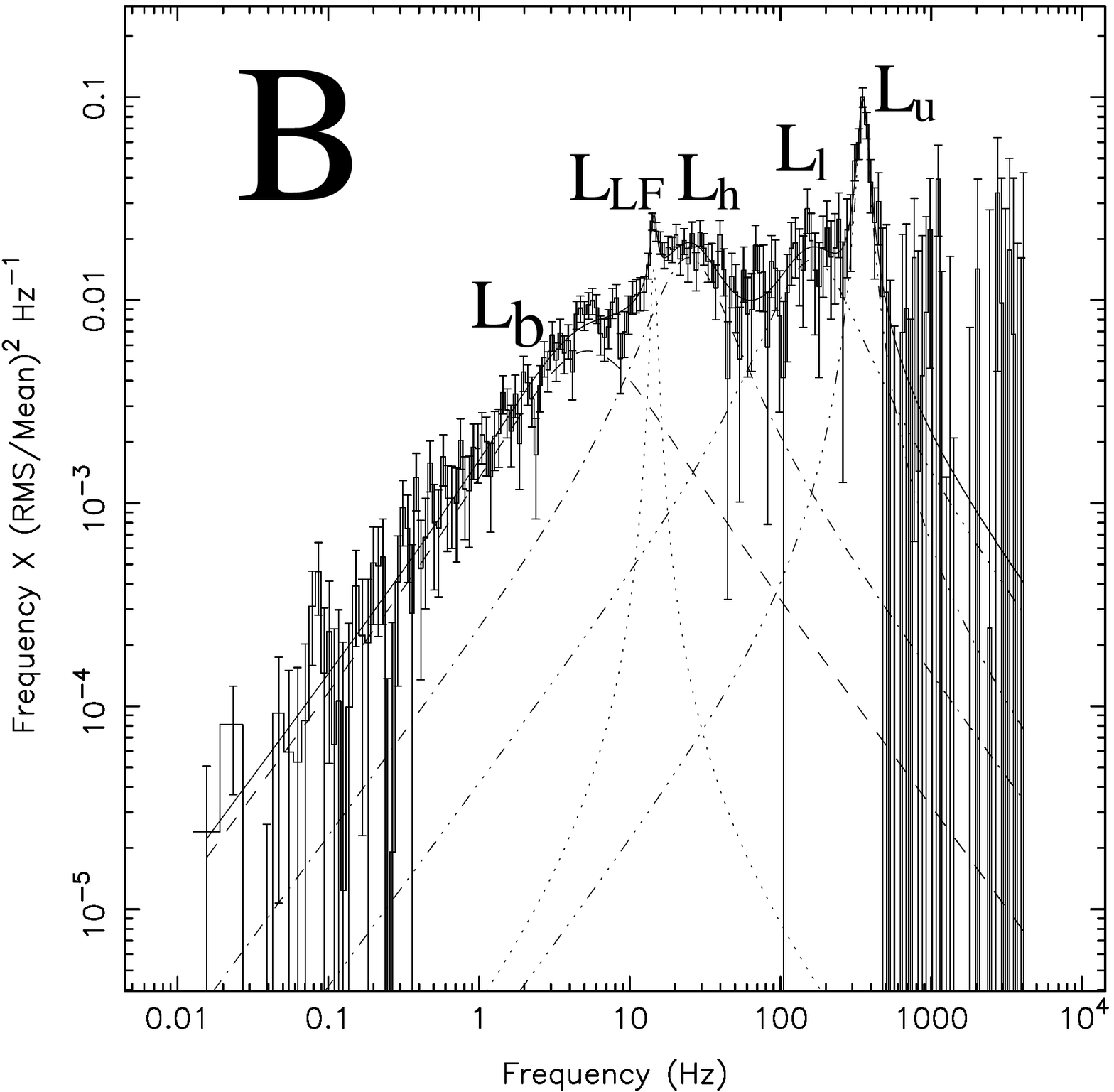}\plotone{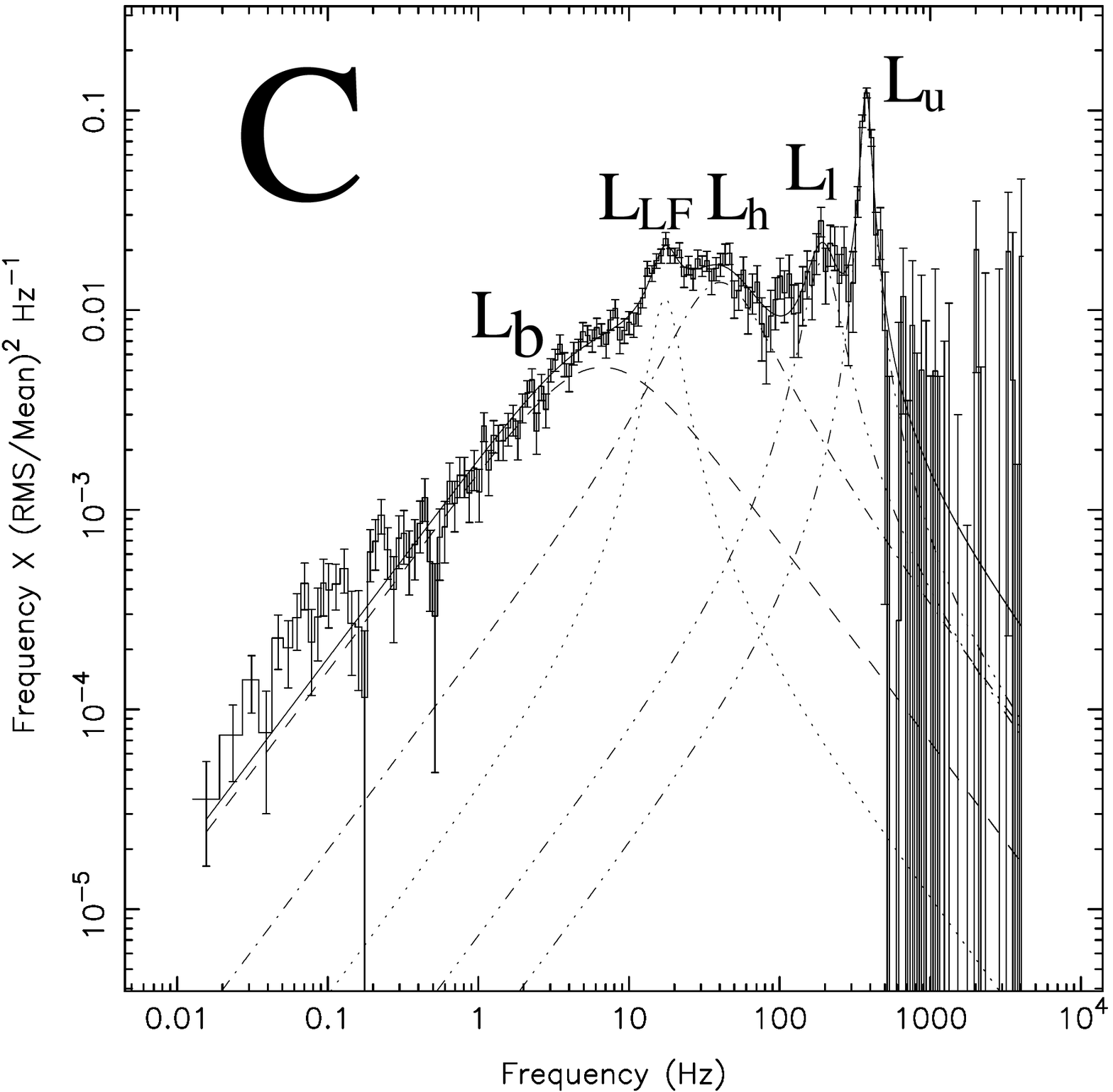}\plotone{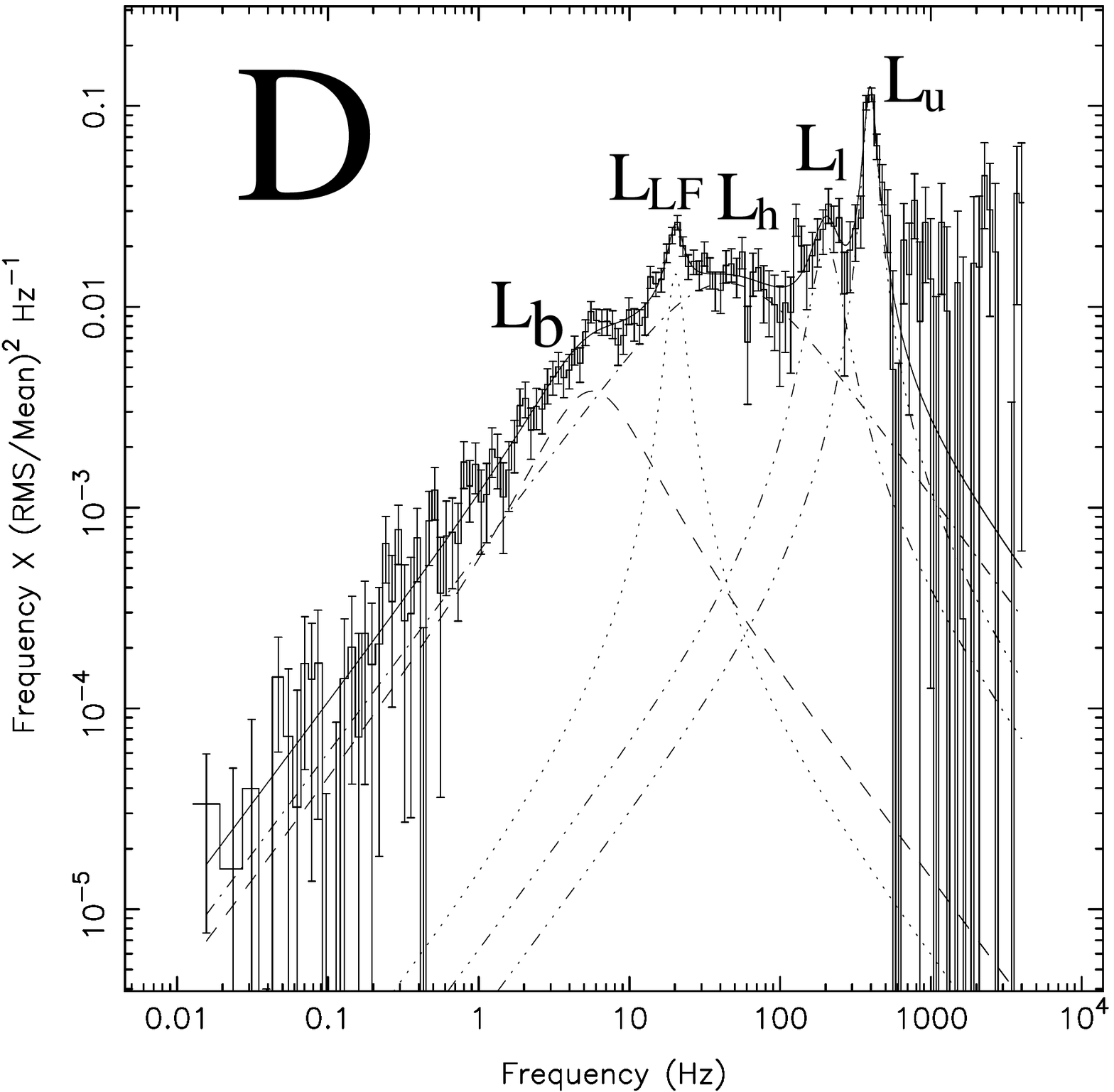}\plotone{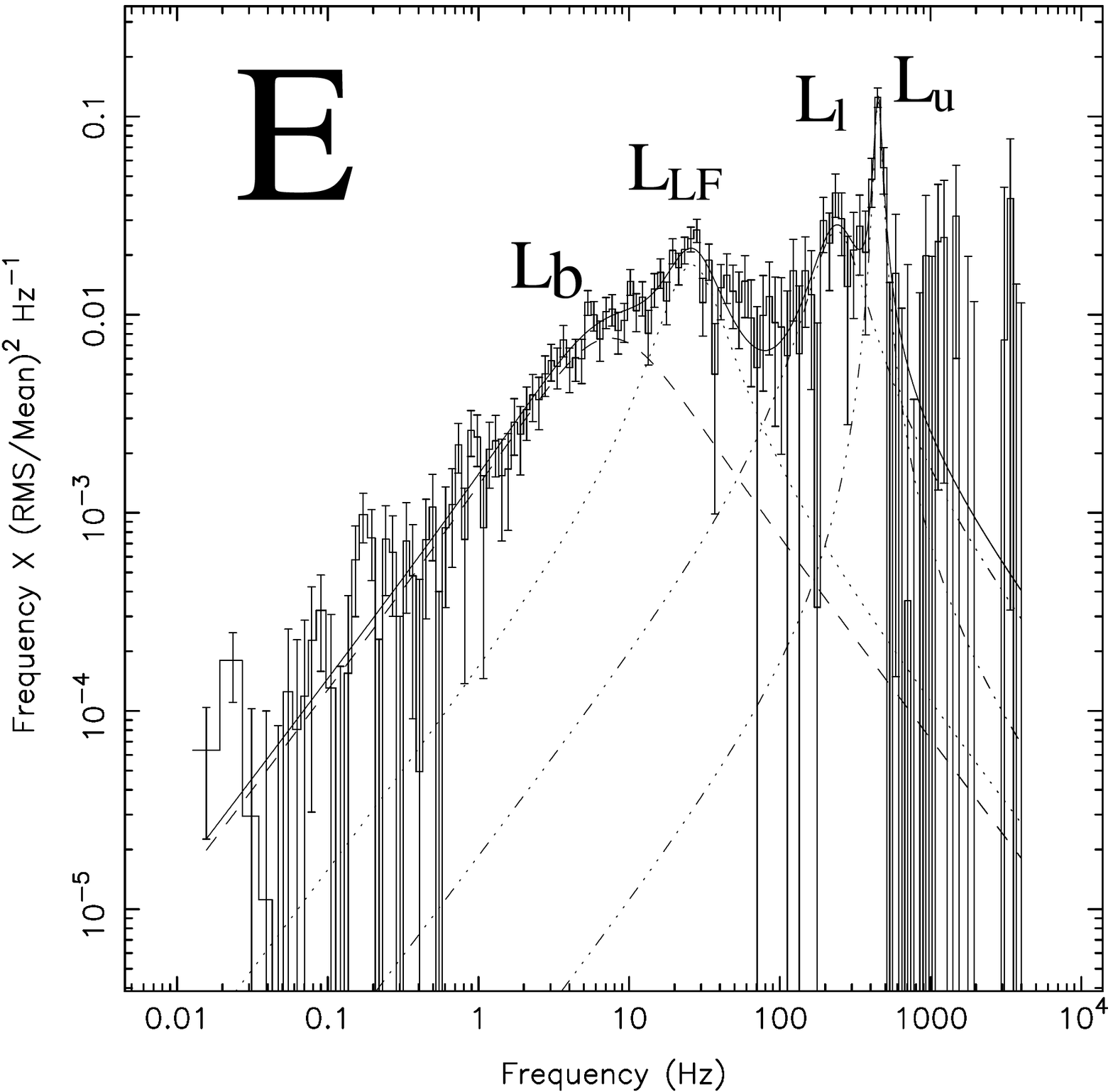}\plotone{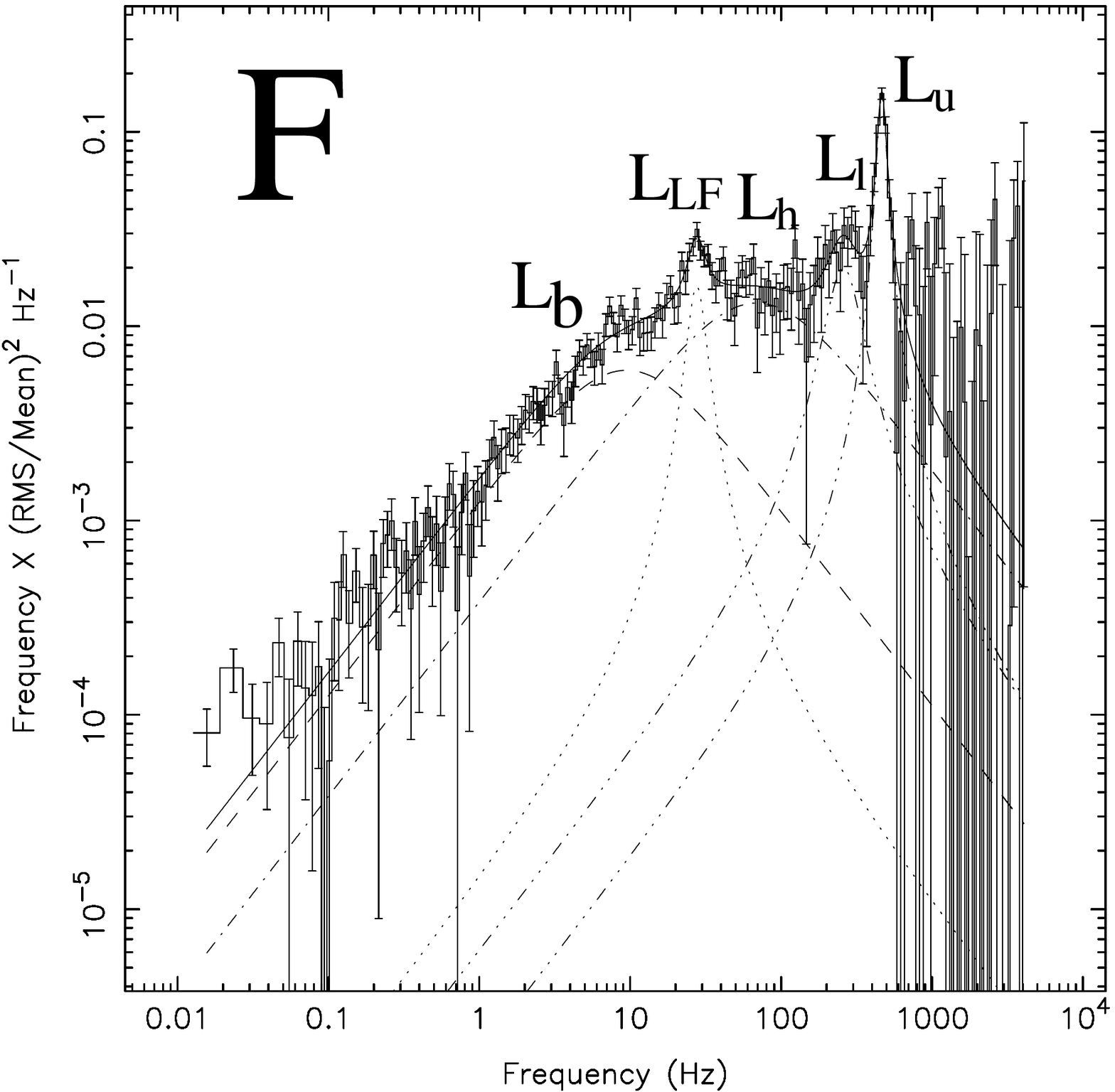}\plotone{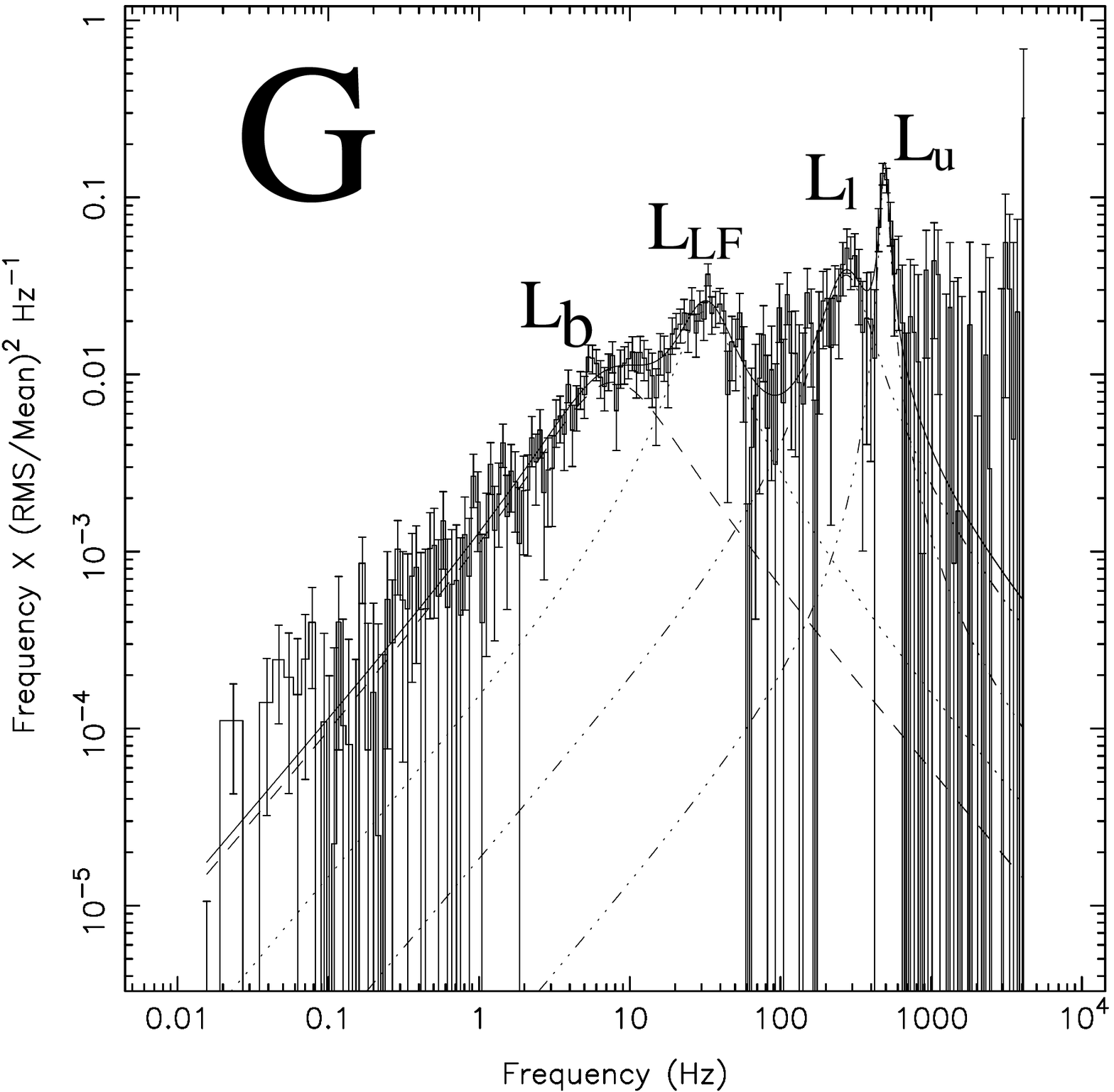}\plotone{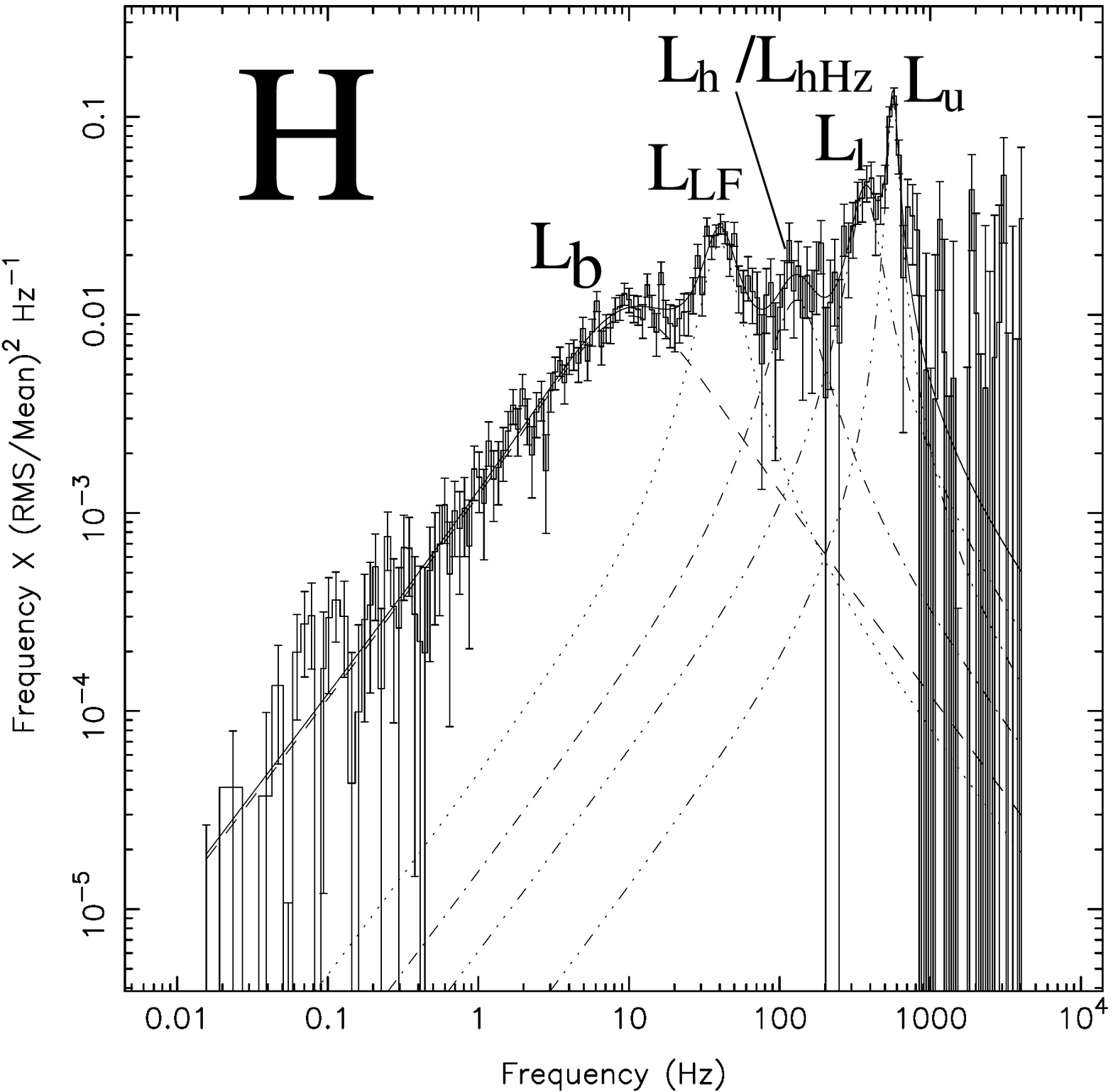}
%\includegraphics[scale=.25]{f6a.eps}%}}
%  \resizebox{0.33\columnwidth}{!}{\rotatebox{0}{
%\includegraphics[scale=.25]{f6b.eps}%}}
%  \resizebox{0.33\columnwidth}{!}{\rotatebox{0}{
%\includegraphics[scale=.25]{f6c.eps}%}}
%  \resizebox{0.33\columnwidth}{!}{\rotatebox{0}{
%\includegraphics[scale=.25]{f6d.eps}%}}
%  \resizebox{0.33\columnwidth}{!}{\rotatebox{0}{
%\includegraphics[scale=.25]{f6e.eps}%}}
%  \resizebox{0.33\columnwidth}{!}{\rotatebox{0}{
%\includegraphics[scale=.25]{f6f.eps}%}}
%  \resizebox{0.33\columnwidth}{!}{\rotatebox{0}{
%\includegraphics[scale=.25]{f6g.eps}%}}
%  \resizebox{0.33\columnwidth}{!}{\rotatebox{0}{
%\includegraphics[scale=.25]{f6h.eps}%}}
  \caption{Power spectra in power$\times$frequency
  vs. frequency representation together with the respective fit
  functions and their Lorentzian components. The pulsar spike was
removed before rebinning in frequency.}
    \label{fig:ps}
\epsscale{1.0}
%  \end{center}
\end{figure}

The broad-band power spectra of the eight sets of observations 
can be seen in Fig.~\ref{fig:ps}, with their 
multilorentzian fits.  The $\chi^2/$d.o.f. 
of the fits were between 155/168 and 192/143. The best-fit parameters 
of the power spectral components are presented in Table~\ref{table:psfits}.

The multilorentzian fits to the power spectra comprise a broad
component at low frequencies (5--10~Hz), $L_b$ (see Section
\ref{sec:data} for nomenclature); two components at intermediate
frequencies, one of them narrow at somewhat lower frequency and the
other one wider and at higher frequency, respectively: $L_{LF}$
(15--40~Hz) and $L_h$ (25--130~Hz); and two narrow components at high
frequencies (the kHz QPOs), $L_{\ell}$ and $L_u$ with characteristic
frequencies 100--370~Hz and 340--570~Hz, respectively.  In two of the
sets (E and G) $L_h$ was not detected, while in set H it has a
frequency of $\sim$130~Hz and could also be identified as $L_{hHz}$
\citep[see][for a description of this component in atoll sources and
AMPs --- the values of the correlation parameters given below do not
change significantly when calculated excluding $\nu_h$ in set
H]{Straaten02,Straaten03,Straaten05}.  The $L_{\ell}$ of the two
lowest frequency sets (A and B) are wider than the rest (Q$\sim$0.5
compared with Q$\sim$ 1.5) and in set A $L_{\ell}$ lies slightly below
the $\nu_{\ell}$ vs. $\nu_u$ correlation (see Fig.~\ref{fig:nunu}),
suggesting that we may be looking at a blend between $L_{\ell}$ and
$L_{hHz}$. In set A we find a narrow Lorentzian (Q=8) at higher
frequency (27~Hz) and a wider one (Q=1.4) at lower frequency
(16~Hz). We call the higher-frequency one $L_h$ and the
lower-frequency one $L_{LF}$; note that this implies $Q_{LF}<Q_h$ for
this data set.  This choice has only a small effect on the numbers
reported below and does not affect our conclusions.  In the sets where
$L_h$ was not detected (E and G), $Q_{LF}$ is also low, which might
indicate that $L_h$ and $L_{LF}$ are blended.

The correlations between our measured characteristic frequencies are
compared to those in other sources in Fig.~\ref{fig:nunu}a.  As in
SAX~J1808.4--3658, our source shows correlations that are shifted with
respect to those of the atoll sources.  Our results are closer to
those of SAX~J1808.4--3658 than to those of the other sources, but are
somewhat different also from that source.  The shift factors between
the correlations in XTE~J1807--294 and those in the atoll sources,
calculated as explained in Section~\ref{sec:data}, are presented in
Table~\ref{table:shifts}.  Note that due to the intrindic dispersion 
present in the frequency-frequency relations the reduced $\chi^2$ ($\chi^2_r=\chi^2/d.o.f.$) is larger than 2 in three of the four shifts presented. We 
take this into account when deriving the errors by using $\Delta\chi^2=\chi^2_r$.  The $L_b$ and $L_h$ correlations show shifts
in $\nu_u$ by factors of 1.46$\pm$0.05 and 1.72$\pm$0.05,
respectively.  The lower kHz QPO can be made to match the atoll
sources by shifting it by another factor ($\sim$1.08) in $\nu_u$ only, but is also
consistent with the option that both $\nu_u$ and $\nu_{\ell}$ are
shifted by the same average factor as derived for $L_b$ and $L_h$ (as
\citealt{Straaten05} reported for SAX~J1808.4--3658). This can be seen in
Fig.~\ref{fig:nunu}b, where $\nu_u$ and $\nu_{\ell}$ of SAX J1808.4--3658
and XTE~J1807--294 have been multiplied respectively by 1.454 (the
value found by \citealt{Straaten05}) and 1.59 (the weighted average of
the $\nu_b$ and $\nu_h$ factors; see Table \ref{table:shifts}). The
figure clearly shows that with these shifts the $\nu_{\ell}$
vs. $\nu_u$ correlation, and for $\nu_u<600$~Hz also the $\nu_b$ and
$\nu_h$ vs. $\nu_u$ correlations, match those in the other sources.

\begin{figure} 
\center
%\resizebox{0.7\columnwidth}{!}{\rotatebox{0}{
\includegraphics[scale=.7]{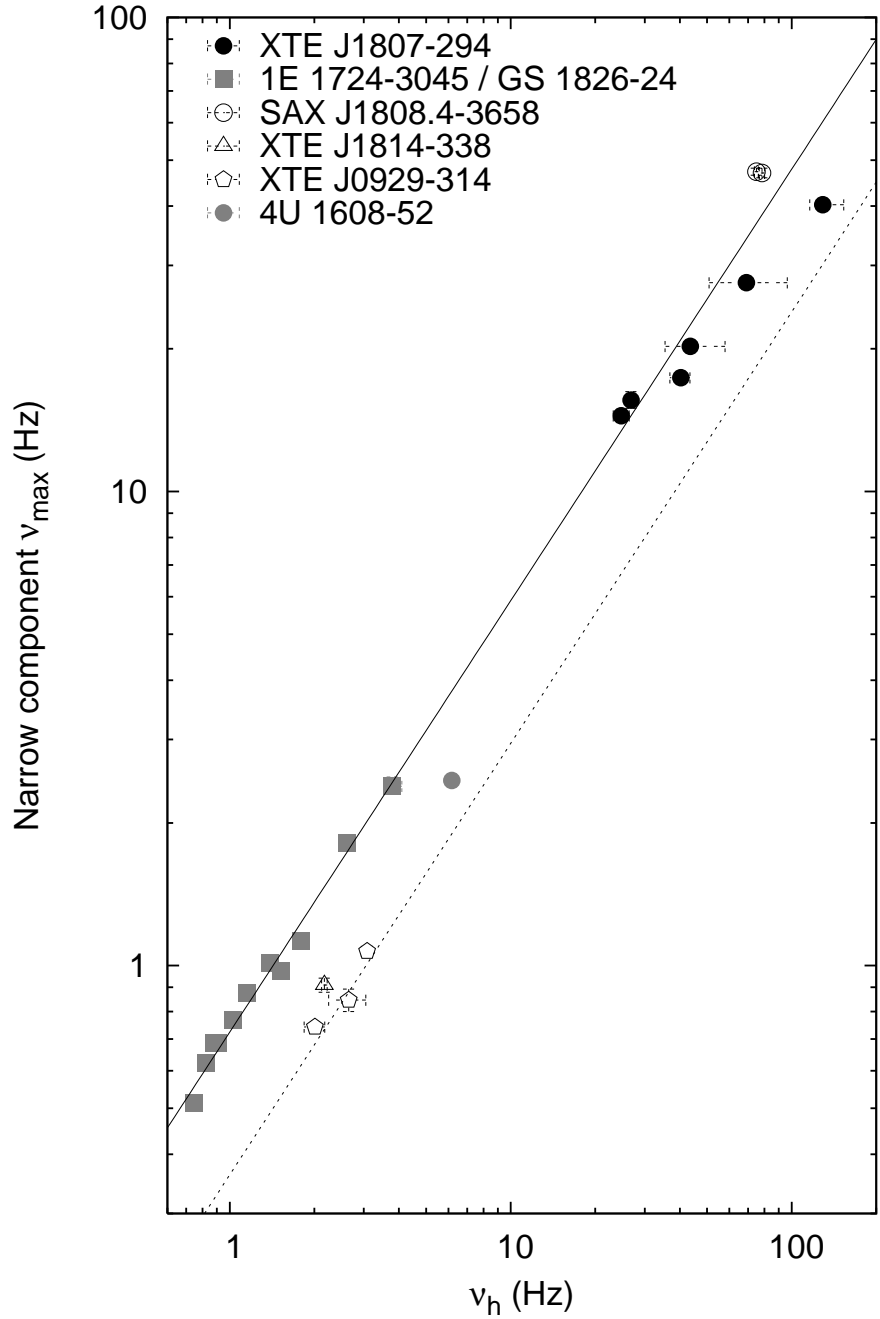}
%}}
\caption{Frequencies of $L_{LF}$ versus the frequencies of $L_h$, for the sources indicated in the plot. The solid line shows the power law fit to the low luminosity bursters 1E~1724--3045 and GS~1826--24. The dotted line represents a power law with the same index and a normalization factor half of that of the solid line.}
\label{fig:nunuLF}
\end{figure}

Following \citet{Straaten05} and previous works, to prevent crowding
of the figures, the $L_{LF}$ components, with frequencies usually
between $\nu_b$ and $\nu_h$, are not displayed in Fig.~\ref{fig:nunu}
but their frequencies are plotted vs. $\nu_h$ separately in
Fig.~\ref{fig:nunuLF}.  The $\nu_{LF}$ values we find in
XTE~J1807--294 are high, above 10~Hz, like in SAX J1808.4--3658
\citep{Straaten05} and in 4U~1820--30 \citep{Altamirano05}, but are
somewhat below the power law extrapolated from the low luminosity
bursters. 

\section{Discussion}
\label{sec:discussion}

We found simultaneous twin kHz QPOs in the power spectra of
XTE~J1807--294. This is the second accreting millisecond pulsar to
show twin kHz QPOs. As the spin period of the neutron star can be
inferred precisely and with high confidence from the pulse frequency
this provides excellent conditions for testing theoretical models of
kHz QPOs (see Section~1; \citealt{vanderKlis04a} for an overview).  In
SAX~J1808.4--3658 the twin kHz QPO centroids were $\sim$195~Hz apart,
consistent with {\it half} the inferred spin frequency
($\nu_{spin}=401$~Hz) and not with $\nu_{spin}$ as the original
beat-frequency models had predicted (see Section~1).  In
XTE~J1807--294, we find instead that the twin kHz QPO separation is near 
{\it once} the inferred spin frequency. $\Delta\nu$ is larger than $\nu_{spin}$
by 2.3$\sigma$. This has previously been seen only in 4U~1636--53 \citep{Jonker02}
and may be difficult to accomodate in beat-frequency \citep{Lamb01} or 
relativistic resonance \citep{Kluzniak04} models.

In subsequent observations of different sources, a picture is
gradually being built up of the relations between the burst
oscillation frequency $\nu_{burst}$, the pulse frequency
$\nu_{pulse}$, the kHz QPO separation frequency $\Delta\nu$, and the
relation of these observed frequencies to the neutron-star spin
frequency $\nu_{spin}$.  We currently have two cases where
$\nu_{burst}\approx\nu_{pulse}$ (and none where these frequencies
differ), three examples of $\Delta\nu\approx\nu_{burst}$ and six
examples of $\Delta\nu\approx\nu_{burst}/2$ (see Table \ref{table:rotators}).  
However, so far the only direct information about
the relation between $\nu_{pulse}$ and $\Delta\nu$ was the one case
(SAX~J1808.4--3658) where $\Delta\nu\approx\nu_{pulse}/2$.  The case
of XTE~J1807--294 constitutes the first example of
$\Delta\nu\approx\nu_{pulse}$.  All these findings are consistent with
the simple picture that (i) the pulse frequency is always the spin
frequency, (ii) the burst oscillations occur always at the spin
frequency, and (iii) for $\nu_{spin}<400$~Hz ('slow rotators'),
$\Delta\nu\approx\nu_{spin}$ while for $\nu_{spin}>400$~Hz ('fast
rotators'), $\Delta\nu\approx\nu_{spin}/2$.  The strict
dichotomy expressed in conclusion (iii) may be due to small sample
size; some overlap between the slow and fast rotator groups might well
be found in further observations. The model of \citet{Lamb03} predicts
that kHz QPOs separated by once and half $\nu_{spin}$ can occur
simultaneously, but this has not yet been observed.  

In SAX~J1808.4--3658, where twin kHz QPOs were detected only once,
several commensurabilities between the centroid frequencies of the kHz
QPOs and the spin frequency of the form $\nu_{j,0}=|m\nu_{k,0} -
n\nu_{spin}|$ (where $\nu_{i,0}$ indicates the centroid frequency of
component $L_i$, $j$ and $k$ designate $u$ and $\ell$ or vice versa
and $m$ and $n$ are integers) were consistent with the data, with in
particular $\nu_{u,0}=|3\nu_{\ell,0}-2\nu_{spin}|$ a remarkably good
match.  We searched for this in XTE~J1807-294, but with seven
detections of kHz QPOs moving over a 200~Hz range in frequency, no
commensurabilities remained consistent with the data as the QPOs
moved.  The data are also not consistent with a constant ratio between
kHz QPO centroid frequencies (Table~\ref{table:qpos}).  This suggests
that, because in SAX~J1808.4--3658 only a single twin kHz QPO pair was detected, 
it was possible for both the commensurability noted
by \citet{Wijnands03} and the 7:4 ratio proposed by \citet{Kluzniak04}
to arise by chance.

\begin{figure}
  \begin{center}
%  \resizebox{1\columnwidth}{!}{\rotatebox{-90}{
\includegraphics[scale=.7,angle=-90]{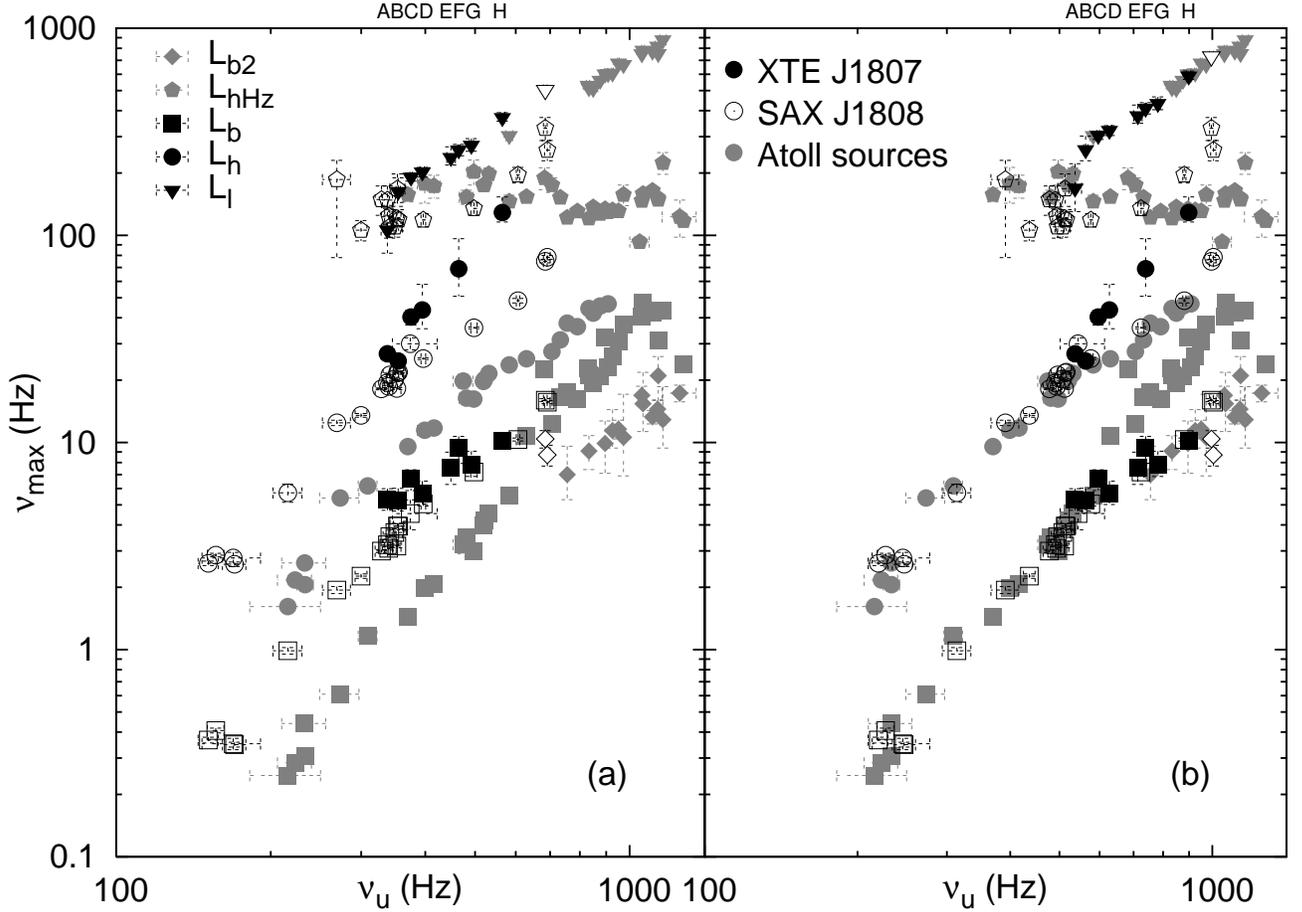}
%}}
  \caption{Characteristic frequencies of the power spectral components
  (except $\nu_{LF}$, which is analyzed separately) plotted versus the
  characteristic frequency of the upper kHz QPO. The black symbols
  correspond to XTE~J1807--294, the open ones to SAX~J1808.4--3658 and
  the grey ones to the atoll sources . Symbols identify the different components as
  indicated, and the letters above the frame the approximate location
  of XTE~J1807--294 data. The observed frequencies are plotted in (a)
  while in (b) the XTE~J1807--294 points were shifted by a factor
  1.59 and the SAX~J1808.4--3658 points by a factor 1.454 in both
$\nu_u$ and $\nu_\ell$. Note the apparent
  bifurcations in the $L_b$ and $L_h$ correlations in (b) for $\nu_u>600$~Hz.}  \label{fig:nunu} \end{center}
\end{figure}

The correlations between the frequencies of the several power spectral
components detected in LMXBs (cf. Section~1) are a relatively new tool to
study these systems, and we are still in the process of figuring out
their significance and even their phenomenology.  The value that we
find for the shift in these correlations ($\sim$1.59) is similar to,
but significantly different from, both the value of \citet{Straaten05}
for SAX~J1808.4--3658 ($\sim$1.45) and the small-integer ratio 3/2=1.5 (a value that could suggest a relation with relativistic resonance models; \citealt{Abramowicz03}).
Yet, the description of the shifts proposed by \citet{Straaten05} as
the most parsimonious one based on the SAX~J1808.4--3658 data (all
$\nu_u$ and $\nu_\ell$ values are too low by the same factor close to
1.5) provides a remarkable match to the data also for XTE~J1807-294
(Fig.~\ref{fig:nunu}b).  In SAX~J1808.4--3658
only a single $\nu_\ell$ value was available on which to base the
conclusion that $\nu_\ell$ was shifted by the same factor as $\nu_u$.
In XTE~J1807-294, blindly applying the same 1.59 shift (derived from the 
$\nu_b$ and $\nu_h$ vs. $\nu_u$ correlations) to the seven well-distributed 
$\nu_\ell$ and $\nu_u$ values,
produces a very good match to the atoll-source $\nu_\ell$ vs. $\nu_u$
relation. 

\begin{figure}
\center
%\resizebox{0.7\columnwidth}{!}{\rotatebox{0}{
\includegraphics[scale=.7]{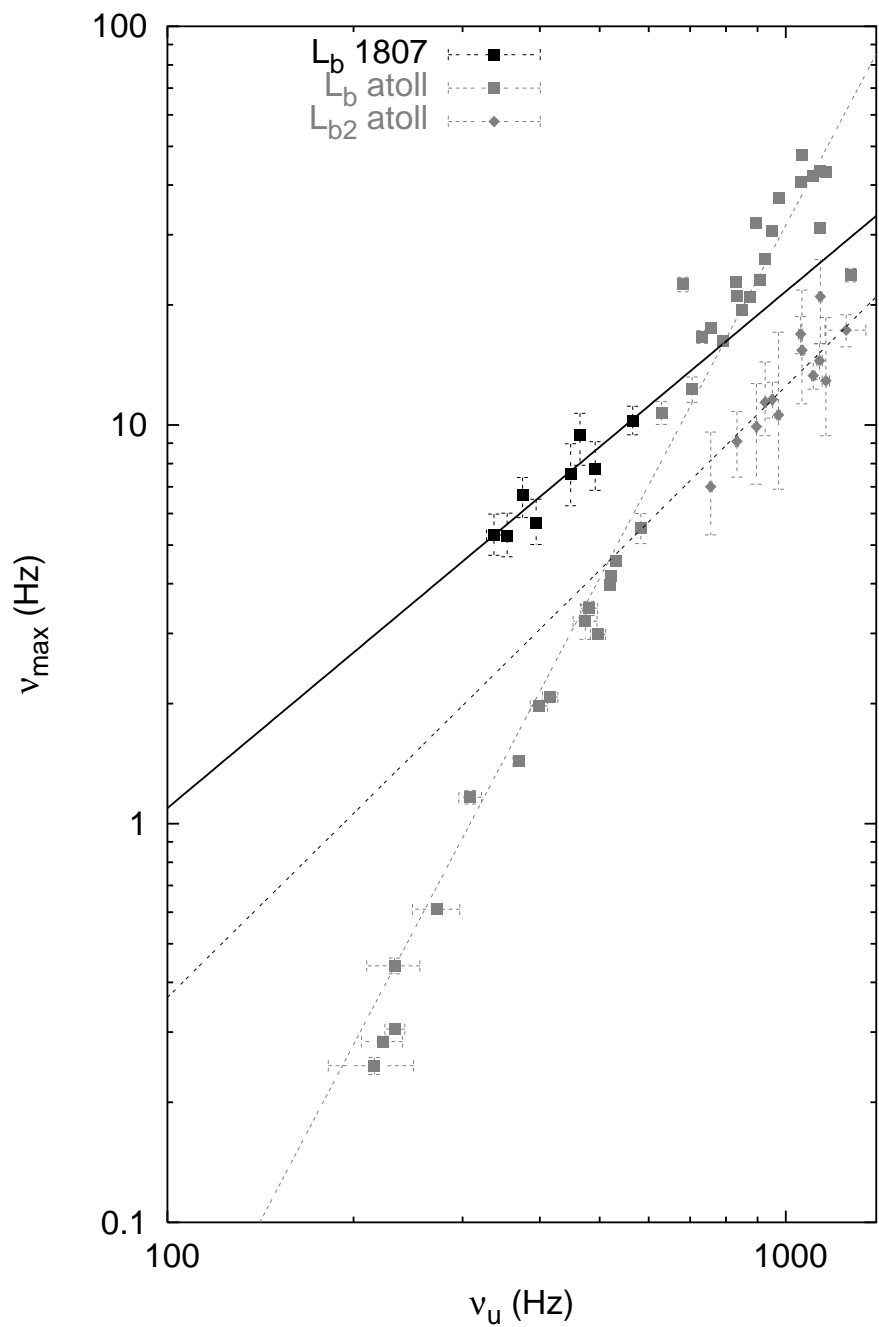}
%}}
\caption{Characteristic frequencies of $L_b$ and $L_{b2}$ for the atoll sources 4U 0614+09, 4U 1608--52, 4U
  1728--34 and Aql X-1 and for XTE~J1807--294, plotted versus $\nu_u$. The fits to the relations are shown as a solid black line (XTE~J1807--294), a dashed black line ($\nu_{b2}$ of the atoll sources) and a dashed grey line ($\nu_b$ of the atoll sources). Five points with doubtful identification were excluded (see Table~\ref{table:powerlaws}).}
\label{fig:nunub}
\end{figure}

\citet{Altamirano05} recently reported a similar
shift in the frequency correlations of the atoll source 4U 1820--30, but by
a factor of $\sim$1.17.  This suggests that the mechanism responsible
for the shifts may also be present in non-pulsing sources and that the
shift factor in $\nu_u$ may be sometimes far from 1.5.
Nevertheless, the small sample size of the 'shifted' sources makes it
too early for final conclusions.  Descriptions involving different
shift factors for different components, and involving shifts of other
components than the kHz QPOs are still possible and, indeed, formally
required by our fit results (see Table~\ref{table:shifts}).  A clear
complication is the nature of the shifted correlation diagram
(Fig.~\ref{fig:nunu}b) at $\nu_u>600$~Hz.  There, the band-limited
noise component labeled $L_b$ in XTE~J1807--294 follows the $L_{b2}$
track of the atoll sources, and the $L_h$ track lies above that for
the atoll sources or SAX~J1808.4--3658.  Power law fits to the $\nu_i$
vs. $\nu_u$ relations of XTE~J1807--294 and the atoll sources
(Table~\ref{table:powerlaws}), show that the slope of the $\nu_b$
vs. $\nu_u$ relation in XTE~J1807--294 is much less steep than in the
other atoll sources, and actually consistent with that of the
$\nu_{b2}$ points (see Fig.~\ref{fig:nunub}), and that the $\nu_h$
points define a {\it steeper} relation in XTE~J1807--294 than in the
other sources. The overall impression created by all of this in
Fig.~\ref{fig:nunu}b is that of a bifurcation in both the $\nu_b$ and
the $\nu_h$ vs. $\nu_u$ relations when $\nu_u$ exceeds 600~Hz.
Perhaps such a bifurcation could be related to the increasing
separation between general-relativistic azimuthal and radial epicyclic
frequencies when approaching the compact object (e.g.,
\citealt{Abramowicz03}), or to the onset of some form of rotational
splitting (cf. \citealt{Titarchuk97}), but at this stage this must
remain conjectural.

Further analysis of other AMPs and atoll sources is needed to gain a
better insight into the frequency correlation shifts. Only a few useful 
power spectra were obtained up to now for XTE~J1751--305, XTE~J0929--314 
and XTE~J1814--338 and no twin kHz QPOs were seen in them. The recent second outburst of XTE~J1751--305 
\citep{Grebenev05, Swank05} or forthcoming outbursts of other new or known AMPs can 
help to give us a more complete view of this shift phenomenon. In any case, our results 
for XTE~J1807--294 when combined with those for SAX J1808.4--3658, strongly 
support a spin-related mechanism for kHz QPO production and strongly suggest 
a spin-independent cause of the shift in the frequency correlations. The
former is due to the first detection of twin kHz QPOs separated by the
spin frequency in a confirmed slow rotator, and the latter to the fact that 
similar shifts are observed in the frequency correlations of at least two AMPs 
with very different spin frequencies.

\textbf{Acknowledgments:}
We thank the referee for valuable comments. ML likes to thank M. Falanga for providing an early copy of his manuscript, and M. Klein-Wolt, T. Maccarone, S.Migliari and R.Wijnands for instructive discussions.
\clearpage

\begin{table}
\center
\caption{Observations used for the timing analysis.}
\begin{minipage}{\textwidth}
%\centering
\begin{tabular}{c c c c c c c}
\hline\hline
Set & ObsID & Start date & Det. \footnote{Minimum and maximum number of active detectors during the observations.} & PDS \footnote{Total number of averaged power density spectra (PDS), each one made from a data segment of 128 seconds.} &  Count rate ($s^{-1}$) \footnote{Time averaged count rate not corrected for the background.} & Bkg. ($s^{-1}$) \footnote{Time averaged background count rate estimated with the standard model for bright sources, version 2.1e, provided by the PCA team at GSFC.} \\ [0.5ex]
 & & & &  & Total -- Per PCU  & Total -- Per PCU \\ [0.5ex]
\hline
 & 80145-01-02-00 & 2003Mar05 & & & & \\ [-1ex]
\raisebox{1.5ex}{A} & 80145-01-04-03 & 2003Mar16 & \raisebox{1.5ex}{3-4} & \raisebox{1.5ex}{126} & \raisebox{1.5ex}{266 -- 70} & \raisebox{1.5ex}{102 -- 27} \\ [1ex]
 & 80145-01-02-03 & 2003Mar06 & & & & \\ [-1ex]
 & 80145-01-02-02 & 2003Mar06 & & & & \\ [-1ex]
\raisebox{1.5ex}{B} & 80145-01-02-01 & 2003Mar06 & \raisebox{1.5ex}{3-5} & \raisebox{1.5ex}{99} & \raisebox{1.5ex}{308 -- 78} & \raisebox{1.5ex}{108 -- 27} \\ [-1ex]
 & 80145-01-03-03 & 2003Mar07 & & & & \\ [1ex]
 & 80145-01-01-03 & 2003Mar01 & & & &\\ [-1ex]
C & 80145-01-01-04 & 2003Mar01 & 4-5 & 104 & 391 -- 91 & 115 -- 27 \\ [-1ex]
 & 80145-01-01-00 & 2003Mar02 & & & & \\ [1ex]
 & 80145-01-01-01 & 2003Feb28 & & & &\\ [-1ex]
\raisebox{1.5ex}{D} & 80145-01-03-02 & 2003Mar07 & \raisebox{1.5ex}{5-5} & \raisebox{1.5ex}{67} & \raisebox{1.5ex}{399 -- 80} & \raisebox{1.5ex}{135 -- 27} \\ [1ex]
E & 80145-01-02-06 & 2003Mar11 & 2-4 & 124 & 215 -- 74 & 84 -- 29\\ [1ex]
 & 70134-09-02-00 & 2003Feb27 & & & & \\ [-1ex]
F & 70134-09-02-01 & 2003Feb27 & 2-5 & 169 & 267 -- 84 & 89 -- 28 \\ [-1ex]
 & 80145-01-02-04 & 2003Mar10 & & & & \\ [1ex]
G & 80145-01-03-01 & 2003Mar12 & 3-4 & 87 & 238 -- 70 & 92 -- 27 \\ [1ex]
 & 80145-01-03-00 & 2003Mar09 & & & &\\ [-1ex]
\raisebox{1.5ex}{H} & 80145-01-02-05 & 2003Mar13 & \raisebox{1.5ex}{2-4} & \raisebox{1.5ex}{154} & \raisebox{1.5ex}{261 -- 77} & \raisebox{1.5ex}{92 -- 27} \\ [1ex]
\hline\hline
\end{tabular}
\end{minipage}
\label{table:obs}
\end{table}
\clearpage
%%%%aqui la taula de resultats dels fits multilorentzians frequencias
%%%%centroidas

\begin{table}
\center
\caption{Twin kHz QPO centroid frequencies with their difference and ratio.}
\begin{minipage}{\textwidth}
\begin{tabular}{c c c c c}
\hline\hline
Set  & $\nu_{u,0}$ (Hz) & $\nu_{\ell,0}$ (Hz) & $\Delta\nu$ \footnote{$\nu_{u,0}$-$\nu_{\ell,0}$} (Hz)  & 
$\nu_{u,0}$/$\nu_{\ell,0}$  \\ [0.5ex]
\hline
B & 352.8$\pm$3.5  & 127.1$\pm$28.0 & 225.7$\pm$28.3 & 2.8$\pm$0.6 \\
C & 374.0$\pm$1.9 &184.0$\pm$10.5 & 190.0$\pm$10.7 & 2.03$\pm$0.12 \\
D & 393.4$\pm$3.1 & 195.0$\pm$14.3 & 198.4$\pm$14.6 & 2.02$\pm$0.15 \\
E & 449.6$\pm$7.1 & 216.5$\pm$23.9 & 233.1$\pm$25.0 & 2.1$\pm$0.2 \\
F & 466.2$\pm$3.0 & 246.3$\pm$19.0 & 220.0$\pm$19.2 & 1.89$\pm$0.15 \\
G & 490.6$\pm$4.6 & 253.9$\pm$22.83 & 236.7$\pm$23.3 & 1.93$\pm$0.17 \\
H & 563.6$\pm$4.6 & 359.9$\pm$15.8 & 203.7$\pm$16.5 & 1.57$\pm$0.07 \\
\hline\hline
\end{tabular}
\end{minipage}
\label{table:qpos}
\end{table}
\clearpage

\begin{table}
\center
%\scriptsize
\footnotesize
\caption{Multilorentzian fits to the power spectra.}
\begin{minipage}{\textwidth}
%\centering
%%
\begin{tabular}{l l c c c c c r}
\hline\hline \\ [-2ex]
  & Parameter & $L_{b}$ & $L_{LF}$ & $L_{h}$ & $L_{\ell}$ & $L_{u}$ & \raisebox{1.5ex}{$\chi^2$} \\ [-1ex]
  &  &  &  &  &  &  &  \raisebox{1.5ex}{d.o.f.} \\ [-1ex]
\hline\hline
  & $\nu_{max}(Hz)$ & 5.3$\pm$0.6 & 15.6$\pm$0.6 & 26.8$\pm$0.5 & 106$\pm$30 & 337$\pm$10 &  \\ [-1ex]
A & Q & 0 (fixed) & 1.4$\pm$0.3 & $8^{+15}_{-3}$ & 0.4$\pm$0.3 & 2.8$\pm$0.6 & \raisebox{1.5ex}{180} \\ [-1ex]
  & rms(\%) & 14.56$\pm$0.8 & 11.6$\pm$1.2 & 5.4$\pm$1.0 & 15$\pm$2 & 16.7$\pm$1.5 & \raisebox{1.5ex}{148} \\ [1ex]
\hline
  & $\nu_{max}(Hz)$ & 5.3$\pm$0.7 & 14.45$\pm$0.17 & 24.7$\pm$1.6 & 163$\pm$23 & 354$\pm$4 &  \\ [-1ex]
B & Q & 0.26$\pm$0.08 & $9^{+11}_{-3}$ & 0.73$\pm$0.18 & 0.6$\pm$0.3 & 5.5$\pm$1.0 & \raisebox{1.5ex}{169} \\ [-1ex]
  & rms(\%) & 11.8$\pm$0.9 & 4.9$\pm$1.0 & 15.9$\pm$1.5 & 16$\pm$2 & 15.8$\pm$1.2 & \raisebox{1.5ex}{165} \\ [1ex]
\hline
  & $\nu_{max}(Hz)$ & 6.7$\pm$0.8 & 17.4$\pm$0.4 & 40$\pm$3 & 191$\pm$9 & 375$\pm$2 &  \\ [-1ex]
C \footnote{An extra component could be added to the fit at $\sim$0.1 Hz, without significant change in the other Lorentzians.} & Q & 0 (fixed) & 2.0$\pm$0.5 & 0.5$\pm$0.2 & 1.7$\pm$0.6 & 6.3$\pm$0.5 & \raisebox{1.5ex}{192} \\ [-1ex]
  & rms(\%) & 12.8$\pm$0.9 & 9.0$\pm$1.3 & 16.2$\pm$1.9 & 12.0$\pm$1.4 & 17.3$\pm$0.5 & \raisebox{1.5ex}{143} \\ [1ex]
\hline
  & $\nu_{max}(Hz)$ & 5.7$\pm$0.7 & 20.2$\pm$0.4 & 44$\pm$12 & 202$\pm$11 & 395$\pm$3 &  \\ [-1ex]
D & Q & 0.45$\pm$0.16 & 3.6$\pm$1.2 & 0 (fixed) & $2.0^{+1.2}_{-0.7}$ & 5.1$\pm$0.6 & \raisebox{1.5ex}{129} \\ [-1ex]
  & rms(\%) & 8.8$\pm$1.4 & 7.8$\pm$1.2 & 20.3$\pm$0.9 & 12.1$\pm$1.9 & 18.9$\pm$0.8 & \raisebox{1.5ex}{121} \\ [1ex]
\hline
  & $\nu_{max}(Hz)$ & 7.5$\pm$1.4 & 26.1$\pm$1.6 & - & 238$\pm$25 & 449$\pm$9 &  \\ [-1ex]
E & Q & 0.2 (fixed) \footnote{Q fixed to 0.2 for the stability of the fit.} & 1.0$\pm$0.5 & - & $1.2^{+1.2}_{-0.6}$ & $7.9^{+4.4}_{-1.5}$ & \raisebox{1.5ex}{132} \\ [-1ex]
  & rms(\%) & 14.2$\pm$1.7 & 15$\pm$3 & - & 17$\pm$3 & 15.2$\pm$1.8 & \raisebox{1.5ex}{111} \\ [1ex]
\hline
  & $\nu_{max}(Hz)$ & 9.4$\pm$1.4 & 27.6$\pm$0.5 & 69$\pm$23 & 259$\pm$16 & 465$\pm$2 &  \\ [-1ex]
F & Q & 0 (fixed) & 3.2$\pm$1.0 & 0 (fixed) & 1.7$\pm$0.7 & 6.7$\pm$0.6 & \raisebox{1.5ex}{202} \\ [-1ex]
  & rms(\%) & 13.6$\pm$1.5 & 8.6$\pm$1.2 & 20.3$\pm$1.1 & 13$\pm$2 & 18.7$\pm$0.7 & \raisebox{1.5ex}{175} \\ [1ex]
\hline
  & $\nu_{max}(Hz)$ & 7.8$\pm$1.1 & 32.1$\pm$1.5 & - & 273$\pm$19 & 492$\pm$5 &  \\ [-1ex]
G & Q & 0.36$\pm$0.11 & 1.0$\pm$0.2 & - & 1.3$\pm$0.6 & 8$\pm$2 & \raisebox{1.5ex}{155} \\ [-1ex]
  & rms(\%) & 14.2$\pm$1.2 & 16.9$\pm$1.4 & - & 20$\pm$2 & 17.3$\pm$1.6 & \raisebox{1.5ex}{168} \\ [1ex]
\hline
  & $\nu_{max}(Hz)$ & 10.2$\pm$0.8 & 40.3$\pm$1.1 & 129$\pm$19 & 370$\pm$18 & 565$\pm$5 &  \\ [-1ex]
H & Q & 0.22$\pm$0.06 & 1.7$\pm$0.3 & $1.2^{+1.4}_{-0.7}$ & 2.0$\pm$0.8 & 6.6$\pm$1.4 & \raisebox{1.5ex}{151} \\ [-1ex]
  & rms(\%) & 16.0$\pm$0.6 & 13.7$\pm$1.1 & $12^{+4}_{-3}$
\footnote{This component was 2.2$\sigma$ single trial but when added
to the fit function it gave a 2.9$\sigma$ improvement of the chi
squared (according to the F-test for additional terms), without
significantly changing the rest of the components.} & 17$\pm$3 &
17.1$\pm$1.4 & \raisebox{1.5ex}{142}  \\ [1ex]
\hline\hline
\end{tabular}
\end{minipage}
\label{table:psfits}
\end{table}

\clearpage

\begin{table}
\center
\caption{Shifts in the frequency-frequency relations in XTE~J1807--294.}
\begin{minipage}{\textwidth}
\begin{tabular}{c c c c}
\hline\hline
Component  & $\chi^2 / d.o.f.$ & $\nu_i$ factor & $\nu_u$ factor \\ [0.5ex]
\hline
$L_{LF}$ \footnote{This shift was done using the $L_{LF1}$ of the low luminosity bursters as a reference (see text and Figure~\ref{fig:nunuLF})} & 22 / 17 & 0.303$\pm$0.031 & 1.725$\pm$0.109 \\
$L_b$    & 68 / 22 & 0.335$\pm$0.030 & 1.462$\pm$0.048$^b$ \\
$L_h$    & 44 / 20 & 0.249$\pm$0.017 & 1.715$\pm$0.048\footnote{The
weighted average of these two values is 1.59$\pm$0.03.} \\
$L_{\ell}$\footnote{A double shift in this relation ($\nu_u$ and $\nu_{\ell}$ equally shifted) gave a factor 1.31$\pm$0.26, with a $\chi^2 / d.o.f.$ of 185 / 21.}    & 185 / 21 & 0.895$\pm$0.083 & 1.082$\pm$0.069 \\
\hline\hline
\end{tabular}
\end{minipage}
\label{table:shifts}
\end{table}

\clearpage

%%%%%%%%%%%%%%%%%%%%%%%%%%%%%%%%%%%%%%%%%%%%%%%%%%%%%%%%%%%%%%
\begin{table}
\center
\caption{Power law fits to the frequency-frequency relations.}
\begin{minipage}{\textwidth}
\begin{tabular}{c c c c c c}
\hline\hline
$\nu_i=N\times{\nu_u}^\alpha$  & $\alpha$ & $\chi^2 / d.o.f.$ & $\nu_i=N\times{\nu_u}^\alpha$  & $\alpha$ & $\chi^2 / d.o.f.$ \\ 
[0.5ex]
\hline
\multicolumn{3}{l}{XTE J1807--294} & \multicolumn{3}{l}{Atoll sources} \\
\hline
$\nu_b$    & 1.29$\pm$0.23 &  4 / 6   & $\nu_b$  & 3.072$\pm$0.093 & 15 / 14 \\
$\nu_{LF}$ & 2.169$\pm$0.070 &  14 / 6   &  $\nu_{b2}$ \footnote{Five $\nu_{b2}$ values measured by \citet{Straaten02, Straaten03} coincide with the $L_b$ track at $\nu_u$  $>$ 600 Hz; these were not included in the measurement of the slope of $\nu_{b2}$ vs. $\nu_u$.} & 1.53$\pm$0.39 &  9 / 11  \\
$\nu_h$    & 3.34$\pm$0.33 &  9 / 4  & $\nu_h$   & 2.530$\pm$0.074 &  29 / 14  \\
$\nu_{\ell}$    & 1.64$\pm$0.14 &   4 / 6 & $\nu_{\ell}$   & 1.402$\pm$0.026 &   181 / 14 \\
\hline\hline
\end{tabular}
\end{minipage}
\label{table:powerlaws}
\end{table}
%%%%%%%%%%%%%%%%%%%%%%%%%%%%%%%%%%%%%%%%%%%%%%%%%%%%%%%

%%%%%%%%%%%%%%%%%%%%%%%%%%%%%%%%%%%%%%%%%%%%%%%%%%%%%%%%%%%%%%
\clearpage
\begin{table}
%\center
\caption{Approximate relations between the kHz QPO separation $\Delta\nu$, $\nu_{burst}$ and $\nu_{pulse}$ in the sources where at least two of these frequencies have been measured. See Table 2.5 in \citealt{vanderKlis04a} for a more extensive version and complete references.}
\vspace{0.3cm}
\begin{minipage}{\textwidth}
\begin{tabular}{l c c c c l}
\hline\hline
 Source  & $\nu_{burst}$ (Hz) & $\nu_{pulse}$ (Hz) & $\Delta\nu/\nu_{burst}$  & $\Delta\nu/\nu_{pulse}$ & \footnotesize{References} \\ 
[0.5ex]
\hline
\multicolumn{5}{c}{Slow rotators} \\
\hline
 XTE~J1807--294 & --- &  191  &   ---   &$\sim$1&\footnotesize{This paper.}\\
 4U~1915--05    & 272 &  ---  & $\sim$1 &  ---  &\footnotesize{\citet{Galloway01}}\\
 XTE~J1814--338 & 314 &  314  &   ---   &  ---  &\footnotesize{\citet{Strohmayer03}}\\
 4U~1702--43    & 330 &  ---  & $\sim$1 &  ---  &\footnotesize{\citet{Markwardt99}}\\
 4U~1728--34    & 363 &  ---  & $\sim$1 &  ---  &\footnotesize{\citet{Mendez99}}\\
\hline
\multicolumn{5}{c}{Fast rotators} \\
\hline
 SAX~J1808.3--3658 & 401 & 401 &$\sim$1/2&$\sim$1/2&\footnotesize{\citet{Wijnands03}}\\
 KS~1731--260      & 524 & --- &$\sim$1/2&  ---  &\footnotesize{\citet{Wijnands97a}}\\
 Aql~X-1           & 549 & --- &$\sim$1/2&  ---  &\footnotesize{\citet{vanderKlis00}}$^a$\\
 4U~1636--53       & 582 & --- &$\sim$1/2&  ---  &\footnotesize{\citet{Wijnands97b}}\\
 SAX~J1750.8--2900 & 601 & --- &$\sim$1/2&  ---  &\footnotesize{\citet{Kaaret02}}\footnote{Marginal detection upper kHz QPO.}\\
 4U~1608--52 & 619 & --- &$\sim$1/2&  ---  &\footnotesize{\citet{Mendez98a, Mendez98b}}\\
\hline\hline
\end{tabular}
\end{minipage}
\label{table:rotators}
\end{table}

%%%%%%%%%%%%%%%%%%%%%%%%%%%%%%%%%%%%%%%%%%%%%%%%%%%%%FIGURES%%%%%%%%%%%%%%%%%%%%%%%%%%%%%%%%%%%%%%%%%%%%%%%%%%%%%%%%%%%%%%%%%%%%%%%%%%%%%%%%

%%%%% figura colors vs time
\clearpage
%% f1.eps

\clearpage
%%%% fig ccd bonic

%% f2.eps

\clearpage

%%%% fig col-col int-col

%% f3.eps

\clearpage
%%%% nice twin kHz QPO IV 

%% f4.eps

\clearpage

%%%% kHz QPO separation

%% f5.eps

%%%% fig power spectra
\clearpage
%% f6.eps
\clearpage

%%%% fig unshifted/shifted nu_max vs nu_u

%% f7.eps

%%%% L_LF in detail:

%% f8.eps

%%%% L_b in detail:

%% f9.eps

%\newpage

\end{document}